\documentclass[twocolumn,english,aps,pra,amssymb,superscriptaddress]{revtex4-2}
\usepackage{mathrsfs}
\usepackage{amsmath} 
\usepackage{hyperref} 
\usepackage{graphicx}
\usepackage{dcolumn}
\usepackage{bm}
\usepackage{natbib}
\usepackage{braket}
\usepackage{circuitikz}
\usepackage{subfigure}
\usepackage{tikz-cd}
\usetikzlibrary{matrix, calc, arrows}
\usepackage{cases}
\usepackage{enumerate}

\begin{document}

\title{Effect of Stochastic Charge Noise in Si/SiGe Quantum-Dot Spin Qubits}

\author{Wei-En Chiu}
\email{b07202014@ntu.edu.tw}
\affiliation{Department of Physics and Center for Theoretical Physics, National Taiwan University, Taipei 106319, Taiwan}

\author{Chia-Hsien Huang}
\affiliation{Department of Physics and Center for Theoretical Physics, National Taiwan University, Taipei 106319, Taiwan}

\author{Yi-Hsien Wu}
\affiliation{Department of Physics and Center for Theoretical Physics, National Taiwan University, Taipei 106319, Taiwan}

\author{Hsi-Sheng Goan}
\email{goan@phys.ntu.edu.tw}
\affiliation{Department of Physics and Center for Theoretical Physics, National Taiwan University, Taipei 106319, Taiwan}
\affiliation{Center for Quantum Science and Engineering, National Taiwan University, Taipei 106319, Taiwan}
\affiliation{Physics Division, National Center for Theoretical Sciences, Taipei 106319, Taiwan}

\date{\today}

\begin{abstract}
 In Si/SiGe quantum dots, the decoherence behavior of spin qubits usually comes from the non-Markovian effect of the charge noise. To improve the performance of using the coherent noise models in the decoherence simulation and tomography analysis, here we propose a spin-phonon model derived from the electric dipole spin resonance to characterize the decoherence behavior of the spin qubit in a Si/SiGe quantum dot. Utilizing a $1/f$ spectrum to characterize quantum noise correlation, our stochastic model can yield a more precise prediction of decoherence compared to a random coherence model. We also use gate set tomography (GST) to address the error generator and analyze the model violation coming from the non-Markovian effect. Based on the results, we attribute certain error generators of this model to the incoherence error, which avoids the scenario of using too large a coherent noise strength in the previous study to account for the experimentally observed decoherence times, and thus underestimates the gate fidelity. We also perform a gate optimization and show that our optimized control pulse can substantially reduce the error contribution of the incoherent non-Markovian $1/f$ charge noise. We further demonstrate that the optimized pulse against incoherent noise is more robust against coherent noise than the regular Gaussian pulse through a filter function analysis in a CPMG protocol, demonstrating the significant effectiveness of the optimized pulse.   
\end{abstract}

\maketitle

\section{\label{sec:level1} Introduction}

Implementing fault-tolerant quantum computation is the ultimate goal in developing quantum computing hardware. Accurately and efficiently implementing quantum gates with reliably high gate fidelity is a prerequisite. Many physical qubit systems, especially solid-state platforms, suffer from non-Markovian noise processes such as $1/f$ noise whose variance diverges with time \cite{PhysRevB.72.134519, Schriefl_2006}. However, the current underlying theories of randomized benchmarking (RB) or gate set tomography (GST) are based on the Markovian approximation \cite{RevModPhys.88.021002, Nielsen2021}. Referring to the memory-less Markovian environment kernel might ignore some important error-accumulation mechanisms. Even employing techniques, such as Pauli twirling \cite{Hashim2023}, to reduce time dependency and noise bias in the characterization results, it is still unclear how to observe accurate non-Markovian system dynamics on the backside. 

There have been several theoretical studies extending quantum tomography methods to adapt to different types of non-Markovian noise. Characterization techniques that produce the most abundant information provided clear evidence that accounts for the non-Markovian effect by comparing the results with experiments \cite{White2020,arxiv.2307.12452, Proctor2020, Mavadia2018, Dehollain2016}. These illustrate the requirement to analyze non-Markovian noise even though related gate characterization protocols cost much more time \cite{li2024}. Therefore, it is still essential to extract the information of non-Markovian noise as much as possible so that one can implement a quantum gate with higher fidelity by suppressing the noise.

In solid-state qubit systems like superconducting qubits or semiconductor Si/SiGe quantum-dot spin qubits, a specially evident non-Markovian noise is $1/f$ charge noise \cite{Yoneda2017, PhysRevB.72.134519, PhysRevB.77.174509, Yoneda2023, arxiv.2309.12542}. The $1/f$ charge noise with a spectrum shape of $1/f$ primarily arises from the electric field fluctuation surrounding the qubit \cite{Keith2022, Dial2016}. A quantum-dot qubit system, where quantum information is encoded in an electron spin in a quantum dot, is influenced by several varieties of non-Markovian environmental noise sources, including local magnetic field inhomogeneity, nuclear spins at isotopes of $^{29}\text{Si}$, and spin-orbit coupling to electric field fluctuations \cite{Yoneda2017, RevModPhys.95.025003, Keith2022, PhysRevLett.88.186802}. For Si/SiGe quantum-dot spin qubits with a micromagnet, electric dipole spin resonance (EDSR) is one of the widely used qubit control techniques to operate the spin qubits \cite{Xue2022, PhysRevLett.96.047202, PioroLadrire2008, PhysRevB.74.165319, PhysRevB.97.085421, PioroLadrire2007,Takeda2016}. Therefore, $1/f$ charge noise can influence the spin qubit through the EDSR mechanism. Although the single-qubit gate fidelity gets higher than 99.9\% and the two-qubit gate fidelity attains 99.8\% in the spin qubit systems \cite{Huang2024, Xue2022, Noiri2022-xw, Mills2022,doi:10.1126/science.aao5965}, obtaining a more detailed description of the non-Markovian noise and developing a control method to suppress the noise is an important step forward in pursuit of even higher fidelities for both single- and two-qubit gates.

Decoherence time is one of the important metrics for evaluating the quality of a qubit. The $1/f$ charge noise plays a role in the decoherence of superconducting qubits \cite{PhysRevB.77.180502}. The $1/f^{\alpha}$ (spin or charge) noise is also believed to be the main factor accounting for the short decoherence time of the spin qubits in Si-based quantum dots \cite{Huang2024, Xue2022, Mills2022, doi:10.1126/science.aao5965, arxiv.2307.12452, Yoneda2023}. Most of the theories that analyzed the decoherence effect of charge noise used a filter function and a two-level ensemble \cite{Schriefl_2006, PhysRevB.72.134519, PhysRevB.77.174509}. Previous characterization protocols demonstrated the acquisition of noise generators with quantum tomography in order to acquire more detailed information on non-Markovian noise \cite{PhysRevA.99.042310, PhysRevApplied.17.024068, arxiv.2307.12452, PRXQuantum.5.010306}. Compared to traditional tomography methods, GST typically requires fewer measurements to achieve a high-fidelity characterization, which is important for practical implementations. Even though we can only get limited aspects of the non-Markovian noise information or extract its error mechanism under the Markovian assumption of GST, it is hopeful to find a noise model that can capture the same characteristics in the GST of the noisy system. From this perspective, we construct a non-Markovian noise model for $1/f$ noise and perform a GST analysis to obtain the error generators, which are consistent with those observed experimentally.

Previous investigations, such as those in \cite{PRXQuantum.5.010306}, employed coherent noise models to simulate the decoherence decay curves of the Ramsey and Carr-Purcell-Meiboom-Gill (CPMG) protocols, yet some deviations from the experimental results remain. Conventional studies typically describe charge noise by introducing an effective classical fluctuating detuning with a $1/f$ spectrum, where the noise strength is derived from the experimentally measured $T_2^*$ \cite{PRXQuantum.5.010306}. Although this phenomenological method successfully reproduces the observed dephasing \cite{Yoneda2017}, it treats the environment as an external classical field and neglects the microscopic quantum interactions responsible for incoherent decoherence. Consequently, relying solely on coherent dephasing to replicate experimental decoherence often demands artificially larger effective noise strengths. Although such models may correctly reproduce the overall decoherence time, they tend to underestimate the achievable gate fidelity and do not fully capture the non-Markovian errors and error distributions observed in gate set tomography (GST) \cite{Yoneda2017}.

To address these limitations and avoid overestimating noise strengths—which artificially suppresses predicted gate fidelities \cite{PRXQuantum.5.010306}—we introduce a comprehensive quantum noise model incorporating non-Markovian $1/f$ charge noise. Our approach utilizes a time-dependent noise channel structured within the master equation framework, where the matrix logarithm of the channel (the "error generator") follows a $1/f$ spectrum. To formulate this model, we revisit the microscopic origin of charge noise arising from Coulomb interactions. Starting from this charge-fluctuation picture, we demonstrate that the conventional classical dephasing term naturally emerges as the leading-order contribution of a multipole expansion, while the next-order dipole interaction gives rise to spin–phonon coupling. This establishes a unified framework where both coherent classical dephasing and incoherent quantum noise originate from the same microscopic mechanism (see Appendix \ref{appendix A}). As a result, our model naturally accounts for an additional incoherent noise channel that enhances the stochastic error generator strength through environmental coupling to the spin qubit.

After describing our noise model for silicon-based spin qubits, as a validation, we simulate the decoherence behavior by performing the same Ramsey experiment numerically to obtain $T^*_2$ and the CPMG sequence for $T_2^{\text{CPMG}}$ with the parameters of a realistic device \cite{Yoneda2017} using a master equation approach. With the established microscopic noise model, we further investigate the GST result to show the usefulness of the error generators and the corresponding average gate fidelity. By writing the master equation and density matrix operators in an extended auxiliary Liouville space, we obtain, using a master equation approach, a gate fidelity closer to the experimental value than in \cite{PRXQuantum.5.010306}. In other words, our model can fit the experimentally measured $T_2^*$ while predicting high consistency with the gate fidelity of the randomized benchmarking (RB) experiment \cite{Yoneda2017}. We also investigate the optimization of the gate operation under this non-Markovian noise model. We adopt the Krotov optimal control method \cite {Krotov1993, Krotov1996} for gate optimization and find that our optimized control pulse can substantially reduce the error contribution of the non-Markovian $1/f$ charge noise.  

The paper is organized as follows. Section \ref{sec:level2} introduces our quantum noise model and its derivation from the charge noise source to the spin-phonon coupling. Section \ref{sec:level3} describes the master equation we use to describe the spin-qubit dynamics for GST and Krotov optimization simulations. It also introduces the advantages of using the Krotov method and presents how to incorporate a non-Markovian noise correlation function into an extended auxiliary Liouville space within the master equation approach for Krotov optimization. Section \ref{sec:level4} shows all our simulation results, including the decoherence simulation of a real qubit device for the validation of our noise model, the addressability of the GST error generators, and the gate optimization result using the Krotov method. The conclusion is given in Section \ref{sec:level5}.

\section{\label{sec:level2} Silicon Spin qubit noise model}

We now introduce the noise model that allows us to connect the error channel to GST. Within the EDSR framework, we explore how charge noise influences the spin-phonon coupling, representing a second-order interaction. Taking into account the correlation times and relative strengths of various noise sources, including charge noise and the nuclear-originated hyperfine interaction, we formulate a decoherence noise model that focuses solely on charge noise.

\subsection{\label{sec:level2.1}  Noise Model Hamiltonian}

In the context of an electron spin qubit within a quantum dot, the electron is typically confined near the interface of the Si/SiGe heterostructure \cite{doi:10.1126/science.aao5965}. In the EDSR scheme, the electron position and spin operators are correlated, which is detailed in \ref{appendix A}, also forming a pathway for charge noise to enter. Ideally, applying an AC voltage to induce oscillations in the electron's position can effectively result in a Rabi oscillation.

The $T_2$ decoherence time in the Si quantum dot is largely affected by $1/f$ charge noise and spin-spin interaction by the nuclear magnetic resonance effect \cite{PhysRevLett.88.186802, Yoneda2017}. The noise carried by the Overhauser magnetic field (hyperfine interaction) can be significantly reduced by using a $^{28}\text{Si}$ isotopically enhanced silicon with a low isotopic concentration of $^{29}\text{Si}$. The correlation time of hyperfine magnetic noise becomes long enough to finish a series of gate operations with the feedback control as shown in \cite{Yang2019-yk}. So, we consider a shorter time regime to specify the charge noise and treat the effect of the hyperfine interaction as a constant in our following analysis of a quantum gate operation \cite{Yang2019-yk, PRXQuantum.5.010306}. Thus, when taking into account the coherence noise on the detuning term, the qubit system Hamiltonian in the rotating frame of Larmor frequency can be cast into a standard form:
\begin{align}
\label{sys_H}
    \hat{H}_S (t) = \frac{\hbar}{2}\delta(t) \sigma_z + \frac{\hbar}{2}\Omega(t) \sigma_x,
\end{align}
where $\Omega(t)$ is our control parameter. Here, $\delta(t)$ contains all frequency detuning parameters, which contains terms that contribute to a decrease in the decoherence time $T_2$ and gate infidelity.
\begin{align}
\label{coherent_noise}
    \delta(t) = \delta_0+ \delta_{\text{c}}\nu(t) + g\mu_{\text{B}}B_n/\hbar,
\end{align}
where $\delta_0$ is the manually tunable frequency detuning, $\delta_{\text{c}}\nu(t)$ is the classical realization of $1/f$ noise and $B_n$ is the effective magnetic field of the hyperfine interaction of the nuclear-spin interaction \cite{PRXQuantum.5.010306}. 

Although some of the devices use a thicker SiGe layer to let the spin qubits of the quantum dots go far away from the oxide layer that contacts the metal gate, because the quantum dots are formed and located near the interface of $\text{Si/SiGe}$ \cite{doi:10.1126/science.aao5965}, there is still some charge-noise effect observed, especially in the decoherence of the spin qubits. It is natural to assume that the fluctuation of the charge densities at the interface contributes to the charge noise. In the rotating frame, the total Hamiltonian of a spin qubit and the environment can be viewed as the Hamiltonian of the spin qubit system combined with second-quantized spin-phonon coupling, whose derivation is outlined in Appendix \ref{appendix A}:
\begin{align}
\label{total_H}
\begin{split}
  \hat{H}&= \frac{\hbar}{2} \delta(t) \sigma_z + \frac{\hbar}{2} \Omega(t) \sigma_x + \sum_k \hbar \omega_k \hat{a}^{\dagger}_k\hat{a}_k \\
  &+ \sum_{k} \hbar g_k \sigma_z  (\hat{a}_{k} + \hat{a}^{\dagger}_{-k}).
\end{split}
\end{align}
 The spin-phonon coupling constant $g_k$ therefore connects the classical charge noise to the quantum noise. We will show that this spin-phonon coupling will also contribute to some of the decoherence in the following master equation construction and the simulation. The inclusion of both classical noise and quantum incoherent noise is also noted in recent developments \cite{Zou2024}.

\section{\label{sec:level3} Master Equation and optimal control in extended Liouville space}

In this section, we use the master equation approach to analyze the error of charge noise and the influence of the spin-phonon coupling of the noise model described by Eq.~(\ref{total_H}). The time convolution of the master equation describes the non-Markovian error even though we use the time-local approximation. Moreover, we can also address the error generators using the master equation approach. We employ an extended auxiliary Liouville space for the master equation to further reduce the calculation complexity, enabling us to use the Krotov optimization method to construct high-fidelity quantum gate operations.

\subsection{\label{sec:level3.1}Time local master equation}
Following the Hamiltonian Eq.~(\ref{total_H}), we now extend the commonly used master equation for qubit dynamics to detailed and applicable equations for qubits influenced by correlated classical and quantum noise. To be more specific, we adopt the time-local master equation approach \cite{PhysRevB.90.104302, Zou2024, 10.1093/acprof:oso/9780199213900.001.0001} to connect both of the classical noise and the quantum noise, leaving the detailed derivation in Appendix \ref{appendix C}. Here, we present time-local master equations for qubits under non-Markovian noise, both in the absence and in the presence of a coherent driving field. This approach enables us to calculate and investigate the impact of non-Markovian noise explicitly. 

The master equation for the reduced density matrix $\rho(t)$ is obtained by tracing out the environment from the total density matrix $\rho_{\rm tot}(t)$, which represents the state of the system and the environment. In the presence of coherent driving, the master equation in the rotating frame of the qubit frequency takes the form (see Appendix \ref{appendix C} for details)
\begin{align}
\label{master_eq}
\begin{split}
    \frac{d\rho(t)}{dt} &= -\frac{i}{\hbar} \left[H_S(t),\rho(t)\right] \\
    &+ \left\{\left[\mathcal{K}(t)\rho(t), \sigma_z\right] + \left[ \sigma_z, \rho(t)\mathcal{K}^{\dagger}(t)\right]\right\},\\
\end{split}
\end{align}
where  
\begin{align}
\label{dissipation_kernel}
    \mathcal{K}(t) = \int^t_0 dt'C(t,t') \mathcal{U}_S(t,t') \sigma_z(t'),
\end{align}
 is the dissipation kernel composed of the bath correlation function 
 \begin{equation}
 \label{bathCF}
   C(t,t')=\sum_k |g_k|^2\braket{\hat{a}_k(t)\hat{a}^{\dagger}_k(t')}  
 \end{equation}
  and the superoperator of the unitary qubit system propagator $\mathcal{U}_S(t,t')$ denoted as
 \begin{align}
 \label{time_ordedring_unitary}
 \mathcal{U}_S(t,t') A= T_+ e^{-\frac{i}{\hbar}\int^{t}_{t'}d \tau [H_S(t), A] }, 
 \end{align}
 with $T_+$ being the time-ordering operator. The $1/f$ charge noise spectrum will specify the bath correlation function $C(t,t')$, and the noise strength is related to the coupling $g_k$. In the time-ordered operator $\mathcal{U}_S$, the superoperator involves the control parameter $\Omega(t)$, thus providing a path to optimize gate operation against the bath-induced non-unitary effect. A similar possibility to revive the coherence of the system against non-Markovian noise was also calculated \cite{Zou2024}.

On the other hand, in the absence of coherent control, the time-local master equation for the qubit system in the frame rotating with the qubit frequency takes the form of
\begin{align}
\label{dephase_no_control}
    &\frac{d\rho(t)}{dt} =  \int^t_0 dt'(C(t,t')+C^*(t,t') )S_Z[\rho(t)],
\end{align}
where $S_P[\rho] = \sigma_{p} \rho\sigma_{p} -\rho$ is the stochastic error generators defined in Appendix \ref{appendix B}. In the rotating frame of the qubit frequency, the qubit loses coherence through the stochastic error generator in the $z$ direction. Without any coherent driving, the nonunitary effect causes information loss from the qubit to the environment, resulting in the qubit being in a mixed state. From the above assumed model, this equation implies that the stochastic error generator is the most dominant error that will be observed in the GST result.

In this dephasing model, we can view the noise terms as an operator 
\begin{equation}
\label{general_noise_op}
\begin{split}
\hat{B}(t)&= \delta(t) \mathcal{I} + \tilde{B}(t)\\
&=\delta(t) \mathcal{I} + \sum_k(g_ke^{i\omega_k t}\hat{a}_k^{\dagger}+g_ke^{-i\omega_k t}\hat{a}_k),
\end{split}
\end{equation}
simultaneously coupling to $\sigma_z$ of the qubit, and $\mathcal{I}$ is the identity operator. The hybrid noise correlation function, which describes the statistical characteristics of classical and quantum bath operators, takes the form of \cite{PhysRevA.95.022121}
\begin{align}
\label{general_corr}
    \braket{\hat{B}(t)\hat{B}(t')} = \braket{\delta(t)\delta(t')}_c + \braket{\tilde{B}(t)\tilde{B}(t')}_q,
\end{align}
where the index $\braket{\dots}_c$ indicates a classical ensemble average, while $\braket{\dots}_q = Tr_B(\dots)$ is a quantum expectation value with respect to the initial bath state. In this case, $\braket{\tilde{B}(t)\tilde{B}(t')}_q$ corresponds to the quantum bath correlation function $C(t,t')$ defined in Eq.~(\ref{dissipation_kernel}) and (\ref{dephase_no_control}). To be consistent with the total noise spectrum observed in the experiment, the classical and quantum noise spectra, Fourier transform of the noise correlation functions, are modeled as a $1/f$ spectrum at low frequencies to simulate the noise effects in the master equation averaged by the quantum and classical ensembles. 

Note that even though the master equation with the noise spectral density of the $1/f$ behavior is time-local \cite{PhysRevA.86.042107}, it is still considered non-Markovian since the dissipation kernel is time-dependent.

The $1/f$ spectrum of the noise could arise from the following scenario. Assume that the correlation function of a noise component decays exponentially $C_{\gamma}(\tau) \propto e^{-\gamma |\tau|}$ with the decay or relaxation rate $\gamma$. Suppose that the distribution of the relaxation rates is proportional to $1/\gamma$, which is usually the case in solid state devices, i.e., $P(\gamma) \propto 1/\gamma$ for a $1/f$ spectrum \cite{Schriefl_2006, RevModPhys.53.497, RevModPhys.86.361, physics/0204033}, and then one can sum up all the correlation functions $C_{\gamma}(\tau)$ for the entire environment and perform a Fourier transform \cite{Schriefl_2006, PhysRevB.77.174509,RevModPhys.86.361, physics/0204033} to illustrate the $1/f$ spectrum as follows:
\begin{align}
\label{1/f_result}
    S(f) = \int^{\infty}_{0}\int^{\infty}_{-\infty} \frac{1}{\gamma} C_{\gamma}(\tau) e^{2\pi i f \tau}  d\tau\, d\gamma \propto \frac{1}{f}.
\end{align}
We will discuss further with respect to how we simulate numerically the $1/f$ noise in a specific frequency range of our interest in the later part of the next section, Sec.\ref{sec:level3.2}.

\subsection{\label{sec:level3.2} Krotov optimization}
Compared to the dynamical-decoupling-based method, optimal control theory is a continuous dynamical modulation with many degrees of freedom to select arbitrary shapes, durations, and strengths for time-dependent control \cite{Tannor1992, PhysRevLett.89.188301, PhysRevA.68.062308, Jirari2009, Wenin2009, PhysRevB.79.224516}. Thus, it allows a significant reduction in the applied control energy and the corresponding quantum gate error, such as gradient ascent pulse engineering(GRAPE) \cite{Yang2019-yk, 9872062} or Krotov optimization methods \cite{Krotov1993, Krotov1996, Tannor1992, PhysRevA.83.053426, Maximov2008, PhysRevA.85.032321, PhysRevA.89.062310, PhysRevA.90.012318, PhysRevA.91.052315}. Moreover, Krotov optimization can provide a more continuous and smoother pulse and can also be implemented in a non-Markovian environment, even for the time-nonlocal equation \cite{PhysRevLett.102.090401, PhysRevA.85.032321}. 

\subsubsection{\label{sec:level3.2.1} Extended Auxiliary Liouville space}
Before executing Krotov's algorithm, we should represent the master equation describing our system in a convenient form for later calculation. We employ an extended auxiliary Liouville space to reduce the calculation complexity, so our correlation function is cast into a discrete numerical form of multi-exponential functions\cite{10.21468/SciPostPhys.7.6.080, PhysRevB.102.035306, PhysRevA.85.032321, Goerz2014-wa,Bartana1993-za, Ohtsuki1999-du,PhysRevA.93.032113}
\begin{align}
\label{numerical_corr_general}
    C(t,t') =\sum_{i} C_i(0)e^{-\theta_i|t-t'|},
\end{align}
where $\theta_i$ and $C_i(0)$ are the numerical fitting parameters which will be selected later for the $1/f$ spectrum. Then the operator Eq.~(\ref{dissipation_kernel}) can be written as 
\begin{align}
\label{sum_kernel}
    \mathcal{K}(t) = \sum_i \mathcal{K}_i(t),
\end{align}
where
\begin{align}
\label{kernel_j}
    \mathcal{K}_i(t) = \int^t_0 dt'C_i(0)e^{-\theta_i|t-t'|} \mathcal{U}_S(t,t') \sigma_z.
\end{align}
This form of $\mathcal{K}_i(t)$ can easily be written in differential equations: 
\begin{align}
\label{kernel_j_diffeq}
    \frac{d\mathcal{K}_i(t)}{dt} = C_i(0)\sigma_z + (\mathcal{L}_S-\theta_i)\mathcal{K}_i(t),
\end{align}
where $\mathcal{L}_S A = -\frac{i}{\hbar}[H_S(t), A]$ with the initial condition $\mathcal{K}_i=0$. With the simultaneous equations Eqs.~(\ref{dissipation_kernel}, \ref{sum_kernel}, \ref{kernel_j}, and \ref{kernel_j_diffeq}), the differential equation Eq.~(\ref{master_eq}) becomes a set of time-local ordinary differential equations without time ordering or memory kernel integration problems. For solving the set of equations numerically, the simultaneous equations can be expressed as the superoperator matrix $\Lambda(t)$ acting on the column vector $\vec{\rho}(t) = [ \rho, \mathcal{K}_1, \cdot\cdot,\mathcal{K}_N]^T$ in an extended auxiliary Liouville space:
\begin{align}
\label{diff_set}
\frac{d \vec{\rho}(t)}{dt} = \Lambda(t) \vec{\rho}(t).
\end{align}
Here, the superoperator matrix consists of the parameters of Eqs.~(\ref{master_eq}) and (\ref{kernel_j_diffeq}). In terms of the propagator $G (t)=G (t,0)$ defined as $\vec{\rho}(t) =G (t,0) \vec{\rho}(0) $, the master equation for iteration will be simplified as 
\begin{align}
\label{Krotov_diff_set}
\frac{d G(t)}{dt} = \Lambda(t) G(t),
\end{align}
 with initial condition $G(0) = I_N$, where $I_N$ is the identity operator.

Next, we describe how we simulate coherent classical noise and incoherent quantum noise, both with $1/f$ noise spectra, in a frequency range of our interest. To numerically simulate the effect of the classical coherent noise with the $1/f$ noise spectrum in a specific frequency interval through individual random realizations, we choose to use the combinations of the multiple  Ornstein-Uhlenbeck processes (OU-processes) to achieve that. The stochastic differential equation of the OU-process $x(t)$ takes the form of 
\begin{align}
\label{OU_process}
    dx(t) = -\gamma\, x(t) dt +\sigma\,  dW(t),
\end{align}
where $\gamma$ is the relaxation rate, or called the mean reversion rate, $\sigma$ is the volatility of the noise, and $W(t)$ denotes the Wiener process. 
The OU-process is a stochastic Gaussian process with the correlation function 
\begin{align}
\label{OU_corr}
    C_{\gamma}(t,t') = \frac{\sigma^2}{2\gamma} e^{-\gamma |t-t'|}.     
\end{align}
The noise spectrum of the OU-process
is a Lorentzian function in frequency 
\begin{align}
    S_{\gamma}(\omega) \propto \frac{\gamma}{\gamma^2+\omega^2}. 
    \label{Lorenzian}
\end{align}
 and by Eq.~(\ref{1/f_result}), 
we can simulate the $1/f$ spectrum in the frequency range in which we are interested using multiple OU-processes with different values of $\gamma$ \cite{Yang2016, physics/0204033}.
Consequently, the combination of noise realizations from these OU-processes can simulate the classical coherent noise $\delta_c v(t)$ in Eq.~(\ref{coherent_noise}) with $1/f$ spectrum:
\begin{align}
    \delta_c v(t)  \propto \sum^{\gamma_\text{cutoff}}  P(\gamma)x(t). 
\end{align}

As the spectral density of the quantum bath (noise) also has a
$1/f$ spectrum in the same frequency range of interest, 
we can use multiple noise modes with different $\gamma$ values, each with an exponential decay form of the correlation function of Eq.~(\ref{OU_corr}), to conveniently construct the bath correlation function of Eq.~(\ref{numerical_corr_general}). By treating the amplitude of each component correlation function as a constant and setting the distribution $P(\gamma) \propto 1/\gamma_i$ for a $1/f$ spectrum, one can then directly determine the parameters in Eq.(\ref{numerical_corr_general}) as 
\begin{align}
\label{theta_i}
    \theta_i=\gamma_i
\end{align}
\begin{align}
\label{C_i}
    C_i (0) = \frac{\sigma_i^2}{2\gamma_i^2}.
\end{align}
The coupling strengths to the bath modes then correspond to the noise strengths in the spectral density via the correlation functions. This construction of the bath correlation function is convenient for us to use in the master equation, Eq.~(\ref{master_eq}), to simulate the effect of quantum noise on the system dynamics. Within the frequency interval of interest in our problem, we select only six modes to reconstruct the bath correlation function for our subsequent simulation.

We briefly discuss the applicability of the hybrid quantum-classical dephasing model with a $1/f$ noise spectral density discussed here for the Si/SiGe quantum dot spin qubits to the Ge/SiGe and silicon metal-oxide-semiconductor (Si-MOS) quantum dot spin qubits.  

Similarly to Si/SiGe quantum dots, Ge/SiGe quantum dot qubits that are confined within a buried strained Ge quantum well sandwiched between crystalline $\text{Si}_{1-x}\text{Ge}_x$ barrier layers are physically shifted $30\text{ nm}$ to $50\text{ nm}$ away from the critical disordered oxide-dielectric interface. Moving away from the oxide interface reduces the absolute magnitude of charge noise reaching the quantum well, but the qubits remain sensitive to defects inside the $\text{Al}_2\text{O}_3$ gate dielectrics, interface steps, and charge traps along the $\text{Si/SiGe}$ heterostructure \cite{Stehouwer2025}. Furthermore, unlike silicon electron qubits, $\text{Ge/SiGe}$ utilizes holes confined in the buried planar germanium quantum well, which occupy the valence band and exhibit an inherently strong intrinsic spin-orbit coupling. The strong SOC implies that the spin states are intrinsically tied to the orbital states and that the qubit has a very high susceptibility to electric-field fluctuations. Charge noise deforms the orbital shape of the quantum dot, leading to severe, direct $g$-factor noise and rapid dephasing \cite{PhysRevB.111.155403}. Thus, in Ge/SiGe quantum-dot spin qubits, the description of dephasing and decoherence centers on the spin-orbit interaction, the anisotropic \(g\)-factor tensor, and \(1/f\) charge noise.

Compared to Si/SiGe qubits, Si-MOS qubits are formed by confining electrons in an inversion layer directly at the amorphous $\text{Si/SiO}_2$ interface. Because the Si-MOS quantum dots sit immediately adjacent to the oxide layer, they experience a high density of near-interface defects and charge traps \cite{arxiv.2306.06229}. Charge noise induces fluctuations in the local electrostatic potential, thereby directly modulating orbital energies and the overlap between adjacent dots \cite{Elsayed2024, Culcer2009}. Thus, two-qubit exchange gates ($\text{CZ}$) may suffer from high infidelities due to noise-induced fluctuations in the inter-dot barrier. Fluctuations in the electric field may alter the valley splitting energy. But if the valley splitting does not become too small, it will not introduce a fast relaxation channel that destroys spin coherence. For single-qubit phase coherence, a Si-MOS spin qubit can remain remarkably resilient to its charge noise environment, as it lacks strong spin-orbit coupling or a micromagnet for qubit control that would reintroduce a pathway through which tiny charge noise can create massive magnetic field fluctuations (causing dephasing). The single-qubit phase coherence of a Si-MOS qubit is predominantly influenced by the randomly fluctuating nuclear spins in the isotope $^{29}\text{Si}$, and this effect is mitigated by isotopically purified \(^{28}\text{Si}\). In isotopically enriched Si-MOS quantum dot spin qubits, reducing the nuclear spin-bearing $^{29}\text{Si}$ isotopes to $50\sim 800$ ppm, the residual spin noise exhibits a characteristic $(1/f)$-like spectral density in the low-frequency regime \cite{Steinacker2025}. In this case, our hybrid quantum-classical dephasing model with a $1/f$ noise spectral density discussed here can be applied to the Si-MOS spin qubit, although the mechanisms causing dephasing may be quite different.

\subsubsection{\label{sec:level3.2.2} Krotov algorithm}

Before describing the Krotov iterative method \cite{Krotov1993}, it is necessary to define an objective function, or a cost function, that we wish to maximize or minimize after each iteration in an optimization algorithm. In realistic quantum gate control problems, it is desirable for the optimal control sequence to achieve the highest gate fidelity with minimal energy consumption.
Therefore, we define the objective function for each iteration $n$ as \cite{Tannor1992, PhysRevLett.89.188301, PhysRevA.68.062308,PhysRevA.85.032321,PhysRevA.90.012318}
\begin{align}
\label{obj_func}
    \mathcal{J}_n = \frac{Tr[Q^{\dagger}G_n(T)]}{N} - \int _0^{T}  dt' \lambda(t) [\Omega_n(t')-\Omega_{n-1}(t')]^2,
\end{align}
where $G_n$ and $Q$ are the actual forward propagator at the final gate time $T$ at iteration $n$ and the target unitary (gate) operator in the extended Liouville space, respectively, and $N$ is the dimension of the matrix $G_n$. $\lambda(t)$ is a positive function that can be adjusted and chosen empirically. The control field energy constraint in the objective function $\mathcal{J}_n$ of Eq.~(\ref{obj_func}), chosen to be the energy in the control pulse difference between the optimized control pulse $\Omega_n(t)$ and the previous control pulse $\Omega_{n-1}(t)$ in the iteration procedure, has the following physical interpretation. The change in control pulse energy in each iteration is limited, and when the iterative procedure approaches the steady optimal solution, the change in the control field is minimal or vanishes \cite{Tannor1992, PhysRevLett.89.188301, PhysRevA.68.062308, PhysRevA.85.032321, PhysRevA.90.012318}. This ensures that the iterative method optimizes the total objective $\mathcal{J}_n$ of Eq.~(\ref{obj_func}) by increasing fidelity rather than reducing the total energy of the control-pulse difference.

We will use the Krotov iterative method \cite{Krotov1993, Krotov1996} to optimize quantum gate control. The Krotov method has several attractive advantages over standard gradient optimization methods:  \cite{Tannor1992, PhysRevA.83.053426, Maximov2008,  PhysRevA.90.012318} (a) a monotonic increase of the objective with the iteration number, (b) no requirement for a line search, and (c) macrosteps at each iteration. In the following, we outline the steps for Krotov optimization in detail \cite{Krotov1993, Krotov1996, Tannor1992, PhysRevA.83.053426, Maximov2008,  PhysRevA.85.032321,  PhysRevA.89.062310, PhysRevA.90.012318, PhysRevA.91.052315}.

\begin{figure}
  \includegraphics[width=\linewidth]{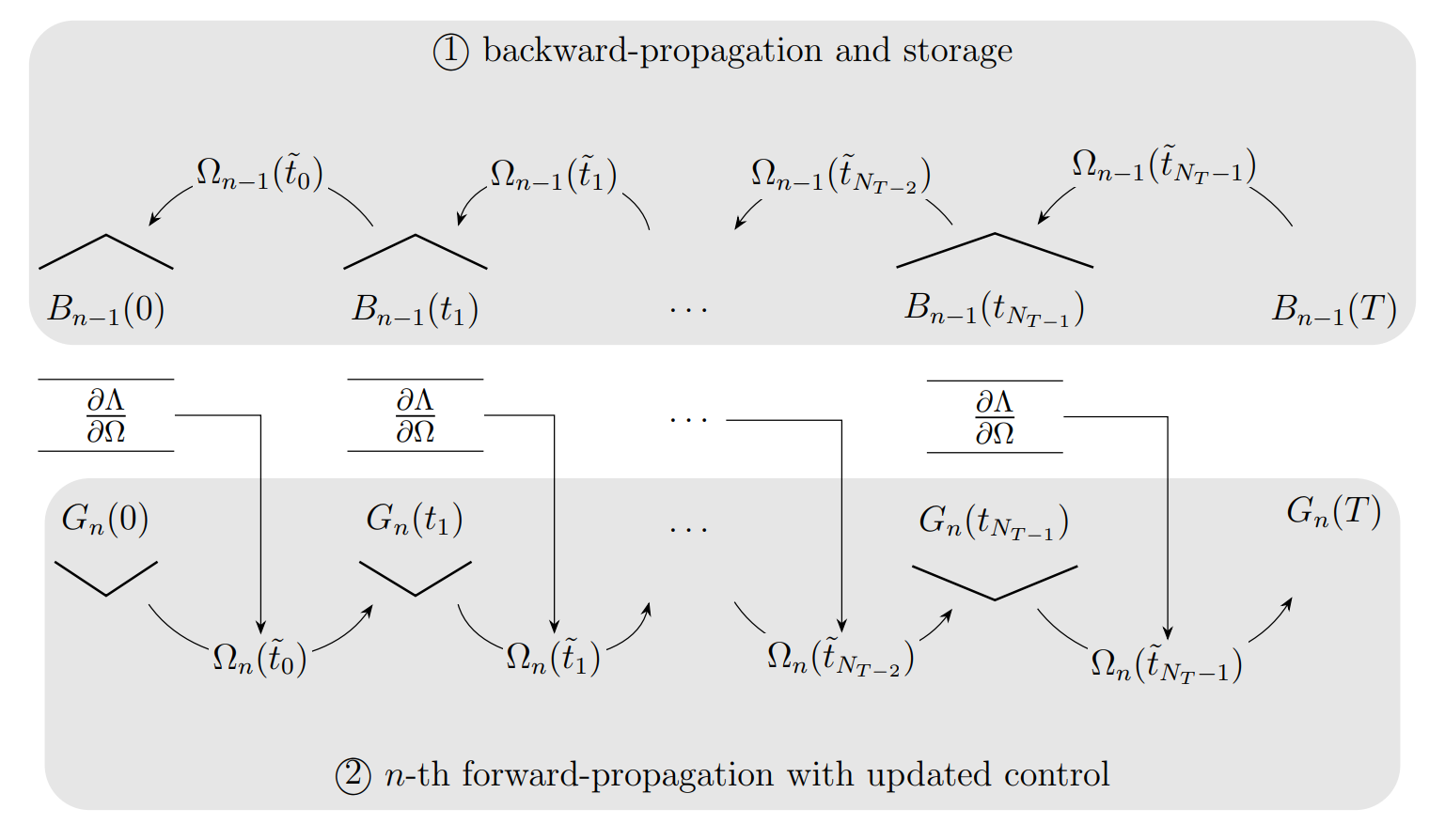}
\caption{\label{iteration_png}  A simple discrete sequential update procedure in one iteration in the Krotov method is illustrated \cite{10.21468/SciPostPhys.7.6.080}. To cut the continuous time evolution into time grids, we label $t_n$ in $G_n(t)$ and $B_n(t)$ in every time grid, and assign $\tilde{t}_n=t_n+dt/2$, where $dt=t_{n+1}-t_n$, for the time grids of control pulses $\Omega_{n}(t)$. In the forward direction with the initial condition $G_{n}(0) = I_N$, the optimized control of the previous iteration $\Omega_{n-1}(t)$, that is, the guessed control of the current iteration, and the backward propagator $B_{n-1}(t)$ of the previous iteration, we can calculate in every time step the control pulse $\Omega_{n}(t)$ from the discretized version of Eq.~(\ref{iteration}) and then the forward propagator $G_n(t)$. After finishing all time steps for $G_n(t)$  and $\Omega_n(t)$ in the $n$-th iteration, we can then calculate the next backward propagator $B_n(t)$ from Eq.~(\ref{backward_eq}) and start the next $(n+1)$-th iteration to find the next forward propagator $G_{n+1}(t)$ and control pulse $\Omega_{n+1}(t)$.}
\end{figure}

\begin{enumerate}[(I).]
\item  Guess a control pulse $\Omega_{0}(t)$ as an initial parameter ($n = 0$ for the first iteration).

\item  Use the equations of motion, Eqs.~(\ref{master_eq}) and  (\ref{Krotov_diff_set}), to find the forward propagator $G_n(t)$ with the initial condition $G_n(0) = I_N$ (beginning with a guess pulse $\Omega_{0}(t)$ for the first iteration $n = 0$ ). 

\item  Find an auxiliary backward propagator $B_n(t)$ with  the equation of motion
\begin{align}
\label{backward_eq}
\frac{d B_n(t)}{dt} =- B_n(t)\Lambda.
\end{align}
and the condition $B_n(T) = Q^{\dagger}$, where $T$ is the (gate) operation time.

\item  Propagate $G_{n+1}(t)$ again forward as time goes on and update the control parameter $\Omega_{n+1}(t)$ iteratively with the iteration rule: 
\begin{align}
\label{iteration}
    \Omega_{n+1}(t) = \Omega_{n}(t)+\frac{1}{2\lambda(t)}Tr\left[ B_n(t)\frac{\partial \Lambda(t)}{\partial \Omega(t)}G_{n+1}(t)\right]. 
\end{align}

Following Eq.~(\ref{iteration}), we present a numerically sequential update scheme in Krotov’s method on a discrete time grid in each iteration to obtain the updated control parameters and the forward propagator in Fig.~\ref{iteration_png} \cite{10.21468/SciPostPhys.7.6.080}.

\item Repeat steps (III) and (IV) until an expected error lower bound is reached or until a given iteration number has been performed.
\end{enumerate}

Note that $1/\lambda(t)$ can be called the step size function, combining the slope produced from its following trace bracket in Eq.~(\ref{iteration}). Thus, Eq.~(\ref{iteration}) can be viewed as a type of gradient descent function, and the step size function,$1/\lambda(t)$, determines each iteration size in Eq.~(\ref{iteration}).

\section{\label{sec:level4} Results and Discussion}

\subsection{\label{sec:level4.1}Benchmarking the noise model with a single spin qubit}

\begin{figure*}
\begin{subfigure}
  \centering
  \includegraphics[width=.3\linewidth]{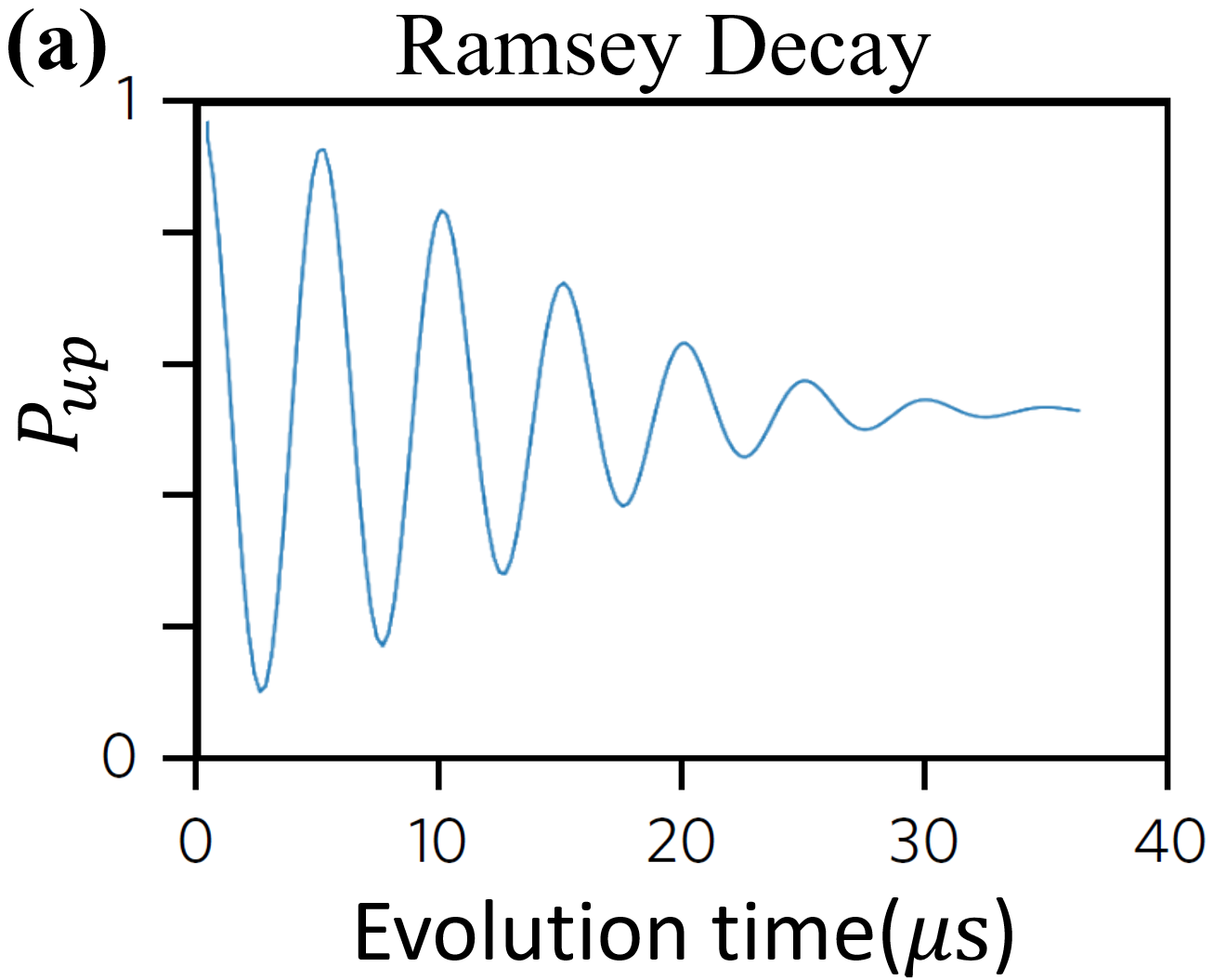}
\end{subfigure}%
\begin{subfigure}
  \centering
  \includegraphics[width=.3\linewidth]{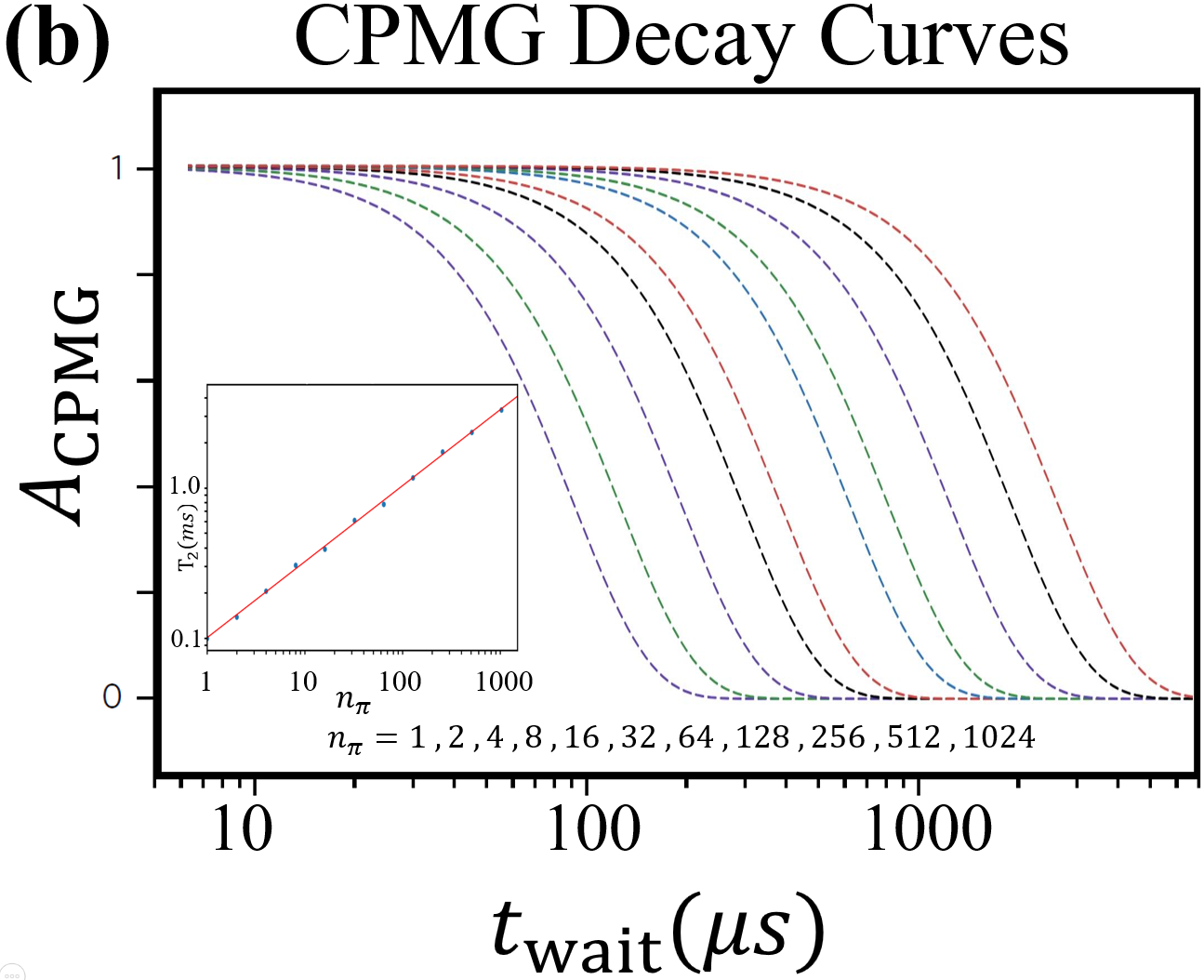}
\end{subfigure}
\subfigure{
  \centering
  \includegraphics[width=.36\linewidth]{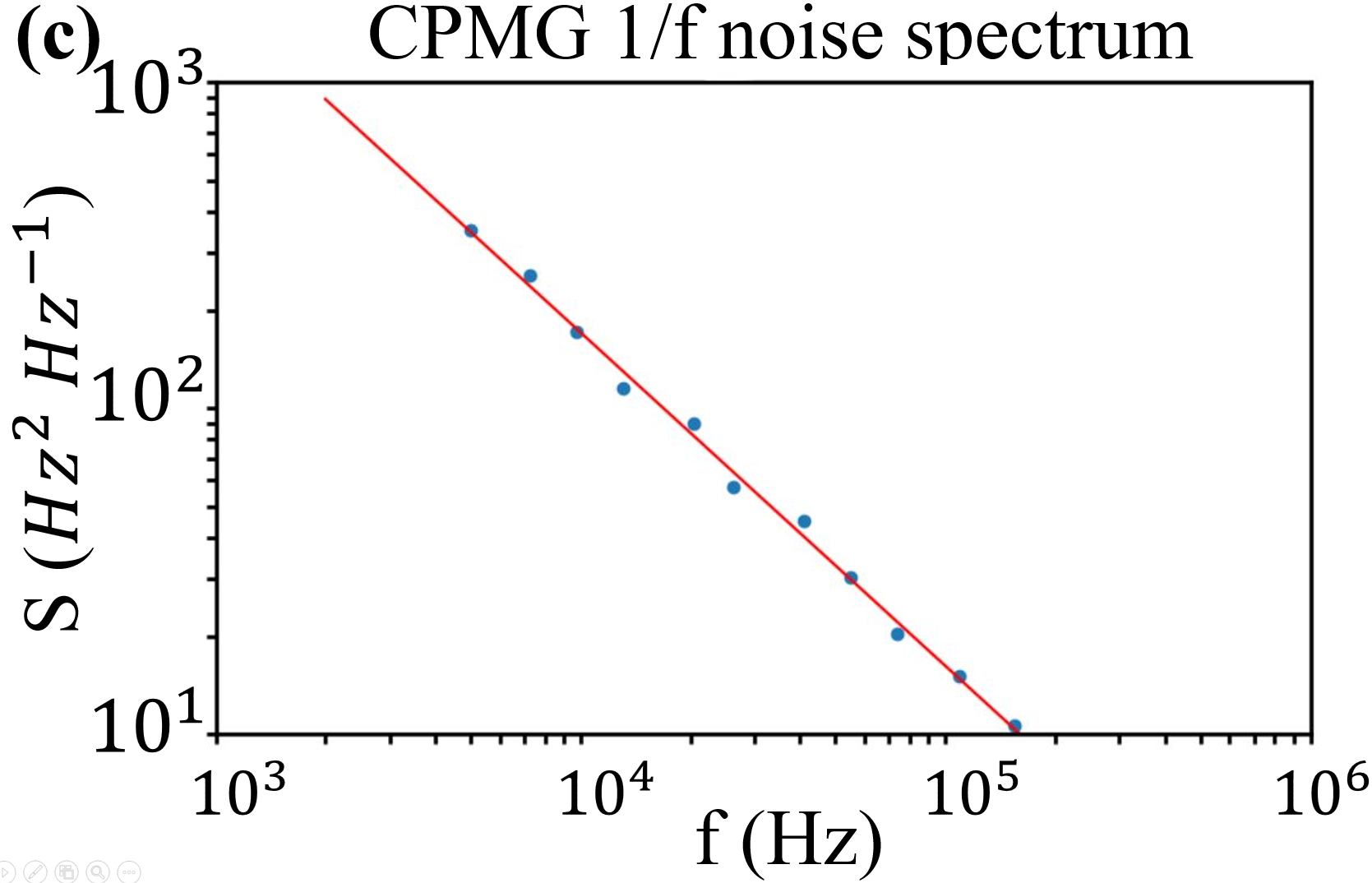}
}
\caption{\label{fig2} We demonstrate numerically simulated decoherence experiments using our model, which includes incoherent noise to better match the real device \cite{Yoneda2017}. (a) The damped oscillation curve (blue line) of the spin-up probability is fitted into the equation $A \cos(2\pi ft) \exp(-(t/T_2^*)^2)+B$, where $T_2^*=24 \, \mathrm{\mu s}$, $f = 194$ kHz, $A=0.5$, $B=0.5$.(b) The dashed lines represent individual decay curves for individual CPMG sequences. Each curve is sorted by the number of $\pi$ gates from left to right. Each line has a constant interval because $T_2^{\text{CPMG}}\propto \sqrt{n_{\pi}}$, as shown in the inset. (c) The $1/f$ spectrum extracted from the CPMG sequence shows a good agreement with the trend line Eq.~(\ref{noise_spectrum}). }
\end{figure*}

In this section, we present results related to our hybrid quantum-classical noise model. Our noise model with the inclusion of quantum noise is compatible with other coherent noise models, which also address $1/f$ charge noise \cite{PRXQuantum.5.010306}. With our model, we can demonstrate a set of gate fidelities closer to the experimental values using parameters extracted from the real device \cite{Yoneda2017}. In addition, we investigate GST features analyzed by error generators to identify possible relations between error generators and operators in the noise model. Within such a model, which includes both incoherent and coherent errors in decoherence, we also optimize the gate operation by reducing the error accumulation rate. The optimized pulse is found to perform better against coherent noise.

 It is important to validate the predictions against established standard noise characterization protocols using the parameters of a realistic device that incorporates the presence of non-Markovian noise processes. The pulse sequences that we employ in our numerical simulations to benchmark the noise model include those in the Ramsey experiment and the CPMG protocol. In these two types of simulations, we demonstrate a similar feature of decoherence and the $1/f$ spectrum relation.

In Fig.~\ref{fig2} (a), we adopt the parameters of one of the real devices in \cite{Yoneda2017} to simulate and observe how close the decoherence behavior of the qubit is to our noise model. These parameters include quasistatic detuning $\delta_{\text{NMR}} \approx 0.2$ MHz in the resonance frequency, charge noise $\delta_c \approx 20$ kHz, for our simulation. In addition, we also use a regular square pulse with the same gate time, $t_{\pi/2} = 15$ ns, to simulate the gate operation. In our simulation, $T_2^*$ denotes the Ramsey decay (decoherence) time when the qubit lies on the azimuthal plane of the Bloch sphere after one Hadamard gate from the spin-up state. To conduct the fitting for $T_2^*$, we first need $\theta_i$ and $C_i(0)$ that appear in the master equation, and then simulate the qubit dynamics in the Ramsey experiment numerically. The values of $\theta_i$ and $C_i(0)$ can be determined by choosing the appropriate values for $\sigma_i$ and $\gamma_i$, for Eqs.~(\ref{theta_i}) and (\ref{C_i}), from the frequency range and suitable noise strength of the $1/f$ noise spectrum for fitting the experimental $T_2^*$ and the average gate fidelity simultaneously. First, a set of $\gamma_i$ can be obtained from fitting the required frequency range (i.e., the frequency cutoff) in the $1/f$ spectrum. After obtaining $\gamma_i$, we determine $\sigma_i$ using Eq.~(\ref{C_i}) by fitting $T_2^*$ and the average gate fidelity simultaneously.  We can use Eq.~(\ref{bathCF}) and Eq.~(\ref{numerical_corr_general}) to connect microscopic coupling to the bath modes with the macroscopic noise strength via the bath correlation function. However, the relevant noise parameters for our simulation are $\gamma_i$ and $C_i(0)$ (or $\sigma_i$). For example, a set of values of $\gamma_i$, $\sigma_i$ and $\sqrt{C_i(0)}$, corresponding to the smallest $\gamma_i$ in our numerical fitting, is approximately 1.0 Hz, 1.7 MHz/$\sqrt{\text{Hz}}$, and 1.2 MHz, respectively. The simulated decay envelope curve, when the qubit loses information and approaches a mixed state, is illustrated in Fig.~\ref{fig2} (a). The decay curve shows $\exp(-(t/T_2)^2)$ at the beginning of the oscillations, demonstrating a non-Markovian decoherence behavior. Our simulation result of $T_2^* \approx 23 \, \mathrm{\mu s}$ yields a very close prediction compared to the result of the reference experiment, $T_2^* = 21 \, \mathrm{\mu s}$ \cite{Yoneda2017}.

Furthermore, our hybrid noise model can give a better prediction of the quantum gate fidelity in the reference experiment \cite{Yoneda2017} than a pure coherence noise model. If only the coherent noise model is used in the simulation, as in \cite{PRXQuantum.5.010306}, the coherent noise strength $\delta_c$ usually needs to be twice as large, i.e, $\delta_c \approx 40$kHz, to get the same $T_2^*$, and therefore the predicted gate fidelity (infidelity) becomes lower (higher). For example, according to the reference device in the experiment \cite{Yoneda2017}, the $Y/2$ gate infidelity is 0.063\%, and the predicted $Y/2$ gate infidelity with only the coherent noise model is 0.126\%, while using our hybrid noise model, the corresponding average gate infidelity is 0.036\%, much closer to the reference value. So, by choosing appropriate noise strength parameters to fit $T_2^*$ in \cite{Yoneda2017}, our noise model introduces another error generator through the dissipation kernel, bridging the coherent noise and the quantum (incoherent) noise, and offers an additional explanation for the qubit decoherence to correctly predict the decoherence time $T_2^*$ and the gate fidelity (infidelity) measured in \cite{Yoneda2017}.

To present the $1/f$ nature of the qubit noise, we conduct the dynamical decoupling simulation in Fig.~\ref{fig2} (b) as in \cite{Yoneda2017}. For the Hahn echo, we apply one $Y$ gate to flip the qubit in the middle of the waiting time when it is on the $x$-$y$ plane, rotating about the $z$ direction to cancel the dephasing effect. In the CPMG protocol, we add various numbers of Y gates to flip the qubit in the waiting time interval. As the number of added $Y$ gates, $n_{\pi}$, increases, the qubit will be considerably decoupled from the random accumulated phase caused by the noise environment. Following \cite{Yoneda2017}, we furthermore relate the normalized echo amplitude $A_{\text{CPMG}}$ to the spectrum $S(f)$ using the filter function formalism for Gaussian noise and then obtain (when $n_\pi \geq 8$) \cite{Yoneda2017, PhysRevB.77.174509}
\begin{align}
\label{noise_spectrum}
    S(\frac{n_{\pi}}{2t_{\text{wait}}})\approx -\frac{\ln (A_{\text{CPMG}})}{2\pi^2t_{\text{wait}}},
\end{align}
and 
\begin{align}
\label{T2_prop_CPMG}
    T_2^{\text{CPMG}}\propto \sqrt{n_{\pi}},
\end{align} 
where $t_{\text{wait}}$ is the total waiting time for one sequence. We present our simulation results for the CPMG protocol in  Figs.~\ref{fig2} (b) and \ref{fig2} (c). The inverse of the frequency in Fig.~\ref{fig2} (c) corresponds to the waiting time $t_{\text{wait}}$, which represents the spectrum fitting in Eq.~(\ref{noise_spectrum}). The non-Markovian error generator does not destroy the $1/f$ noise spectrum revealed in the CPMG protocol. It can be explained that classical noise and quantum noise are effectively reduced by addressing the corresponding noise strength in the spectrum through frequency analysis. Increasing $n_{\pi}$ statistically reduces noise and coupling.  We will utilize the filter function formalism in the CPMG protocol to demonstrate that the optimized gate, which mitigates quantum noise with a $1/f$ spectrum, can also significantly reduce the effect of classical noise with a similar $1/f$ spectrum in Sec.~\ref{sec:level4.3}.

Incorporating the overall effect of an error generator of a non-Markovian incoherent dephasing noise, Eq.~(\ref{dephase_no_control}), the off-diagonal decoherence in the Ramsey simulation can be predicted approximated as (see Appendix \ref{appendix D})
\begin{align}
\label{T2_Ramsey_decay}
    \rho_{01} = \frac{1}{2} e^{-\tilde{G}(t)},
\end{align}
\begin{align}
    \tilde{G}(t) = \sum_i \frac{\sigma_i^2}{\gamma_i^3} \left( t+\frac{1}{\gamma_i}e^{-\gamma_i t}-\frac{1}{\gamma_i} \right),
\end{align}
where  $\tilde{G}(t)$ is derived in Appendix \ref{appendix D}. By expanding the series of $\tilde{G}(t)$ in time, we can have an approximation of Eq.~(\ref{T2_Ramsey_decay}) as
\begin{align}
\label{T2_approx}
    \rho_{01}\approx \frac{1}{2}e^{-(\frac{t}{T_2})^2},
\end{align}
which is consistent with the observed Gaussian decay behavior in our simulation in Fig.~\ref{fig2} (a) and the experiment in \cite{Yoneda2017}. Then combining the Hahn echo and CPMG results of Eq.~(\ref{T2_prop_CPMG}) and our decoherence function Eq.~(\ref{T2_approx}), one obtains
\begin{align}
\label{T2_decay}
    \rho_{01} = \frac{1}{2} e^{-\frac{\tilde{G}(t)}{\mathnormal{m} n_{\pi}} },
\end{align}
where $m$ is the constant determined by the Hahn echo simulation and parameter fitting, thus showing that the result in Fig.~\ref{fig2} is close to that in \cite{Yoneda2017}. This equation demonstrates that the quantum noise model offers greater flexibility in parameterizing the nature of qubit noise.

\subsection{\label{sec:level4.2}Gate set tomography and gate infidelity}

To fully investigate the error source and its channel, it is common to use gate set tomography (GST) to analyze all components of an operation. We leave some of the basic properties and detailed analysis of error generators in GST in Appendix \ref{appendix B}. In GST, the coefficients of the elementary error generators are represented in the Pauli basis with the Pauli transfer representation; thus, the error generators acting on an initial state $\rho_0$ through a sequence of gate operations $\mathcal{G}_1$ and $\mathcal{G}_2$, can be interleaved between each gate with a matrix product as follows:
\begin{align}
    \mathcal{G}_2 \circ \mathcal{G}_1[\rho_0] \rightarrow e^{\mathcal{L}_2} G_2 e^{\mathcal{L}_1} G_1 \vec{\rho}_0.
\end{align}
Here, for a single qubit, $\vec{\rho}_0$ is projected from a 2-dimensional matrix into a 4-dimensional vector in a vector space expanded by Pauli operators, and $G_1, G_2$, $e^{\mathcal{L}_1}, e^{\mathcal{L}_2} $ are 4-dimensional matrices transformed from the operators of quantum channels $U[\cdot]=U\cdot U^{\dagger}$, $\mathcal{E}[\rho]=\rho+\mathbb{L}[\rho]$ in Eq.~(\ref{quantum channel}) and Eq.~(\ref{error generator}). Therefore, an operation can be decomposed into one error generator and one ideal gate matrix, and their operation sequence is also simplified by the matrix product. When we repeat the germs, consisting of selected gates, to magnify the error in the gate, it is convenient to observe the accumulation of errors in the error generators. 

GST provides rich information on quantum operations with a convenient package pyGSTi \cite{https://doi.org/10.11578/dc.20190722.2}. Most of the research characterizing the non-Markovian effect in quantum gates focuses on error generators, including their time variation or spectrum, using various tomography methods \cite{PRXQuantum.5.010306, arxiv.2307.12452, PhysRevApplied.17.024068}. Based on the results of the error generators that vary at a slow frequency, we can observe that the dominant ratio does not change significantly. Here, we investigate the relation between the taxonomy of error generators and our quantum noise model using GST. Some detailed analysis of error generators in GST is presented in Appendix \ref{appendix B}.

\begin{figure}
\begin{subfigure}
  \centering
  \includegraphics[width=.48\linewidth]{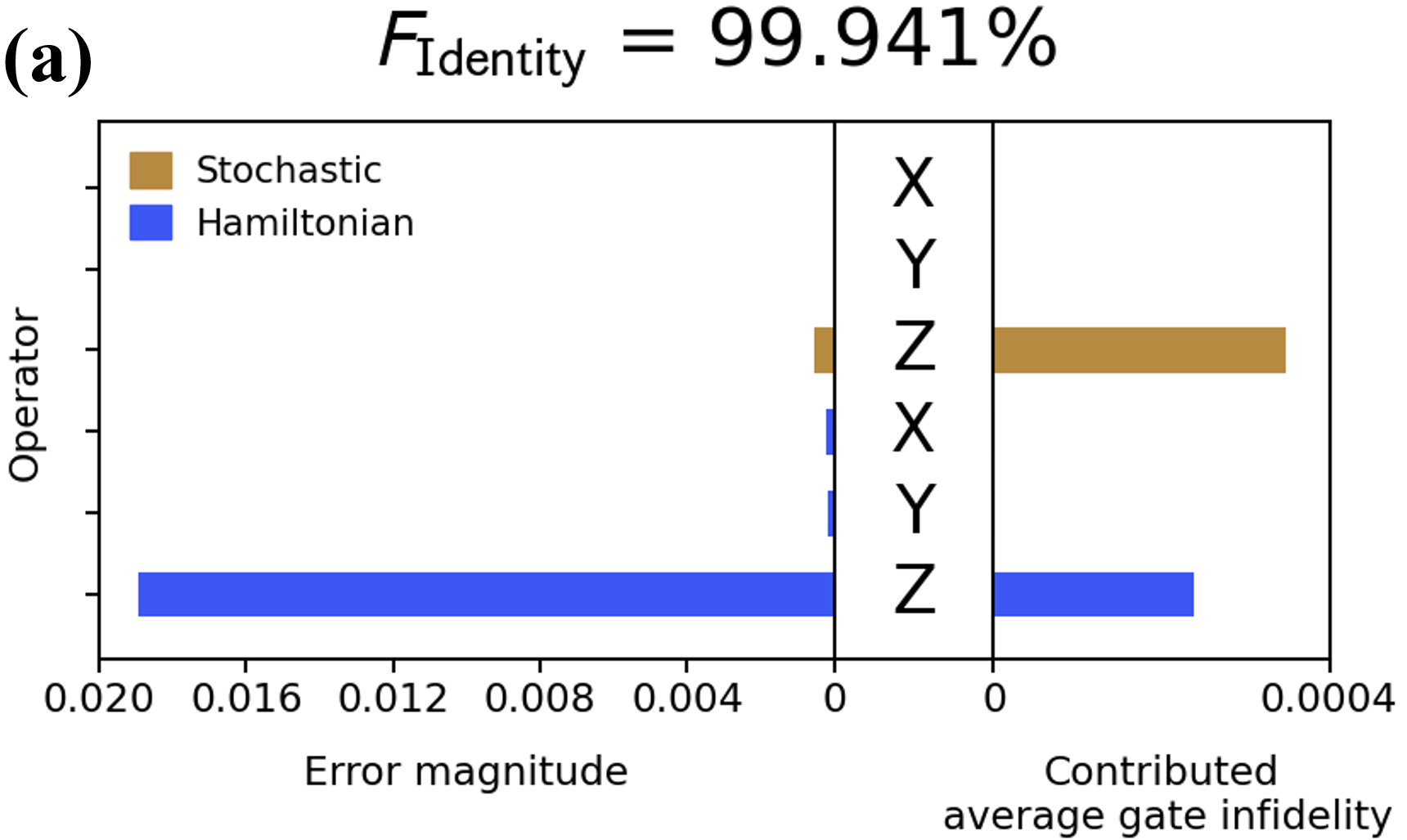}
\end{subfigure}%
\begin{subfigure}
  \centering
  \includegraphics[width=.48\linewidth]{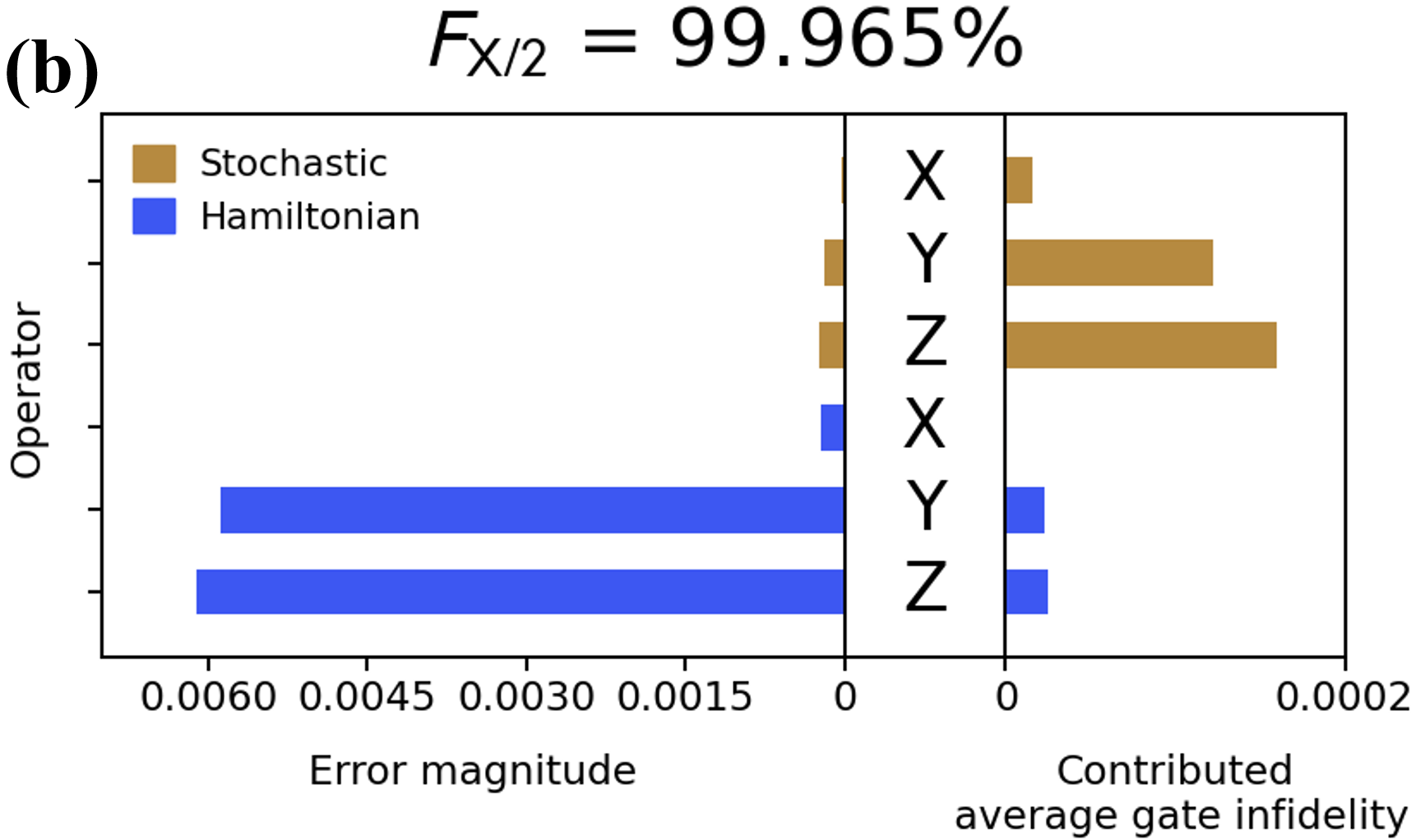}
\end{subfigure}
\begin{subfigure}
  \centering
  \includegraphics[width=.48\linewidth]{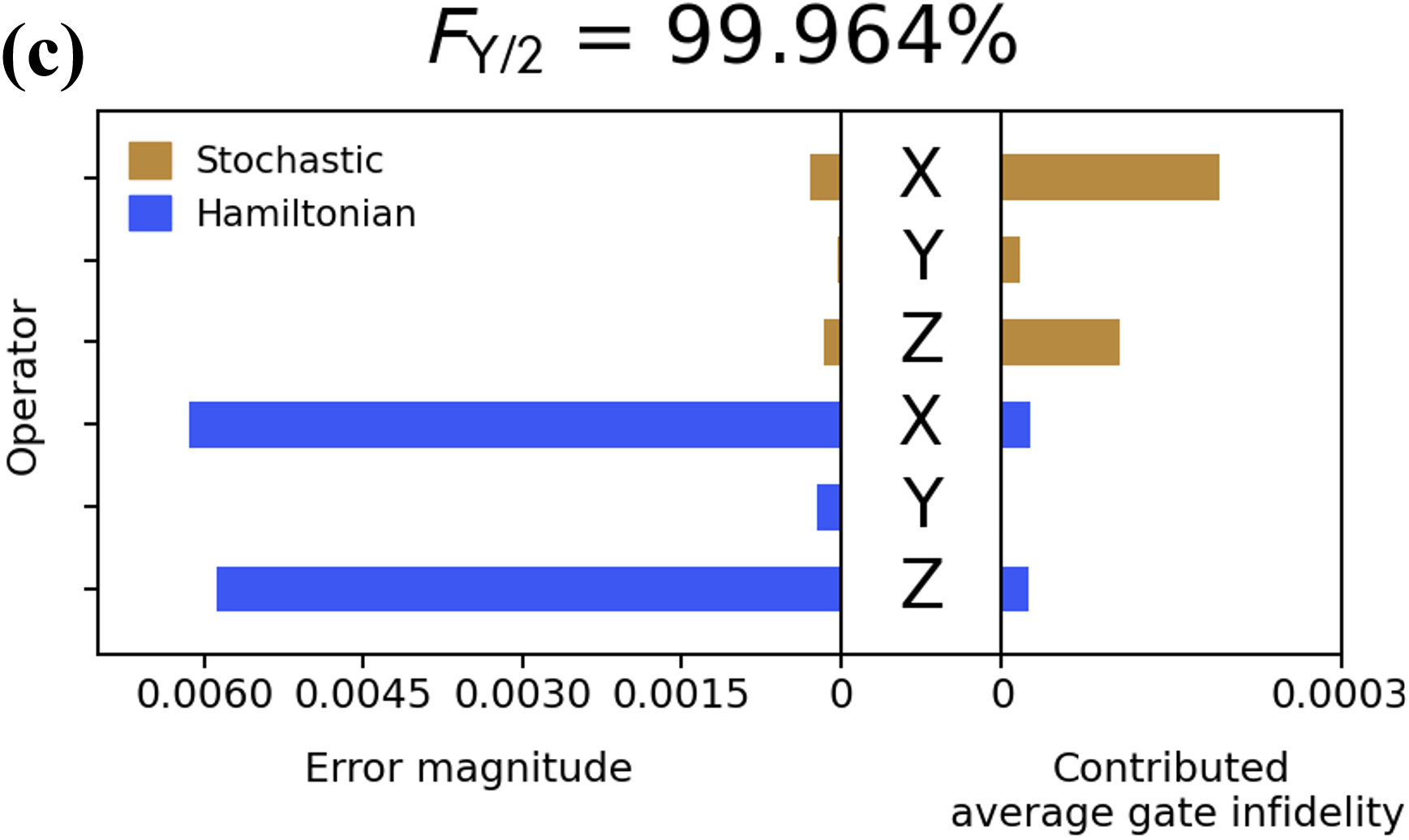}
\end{subfigure}
\begin{subfigure}
  \centering
  \includegraphics[width=.48\linewidth]{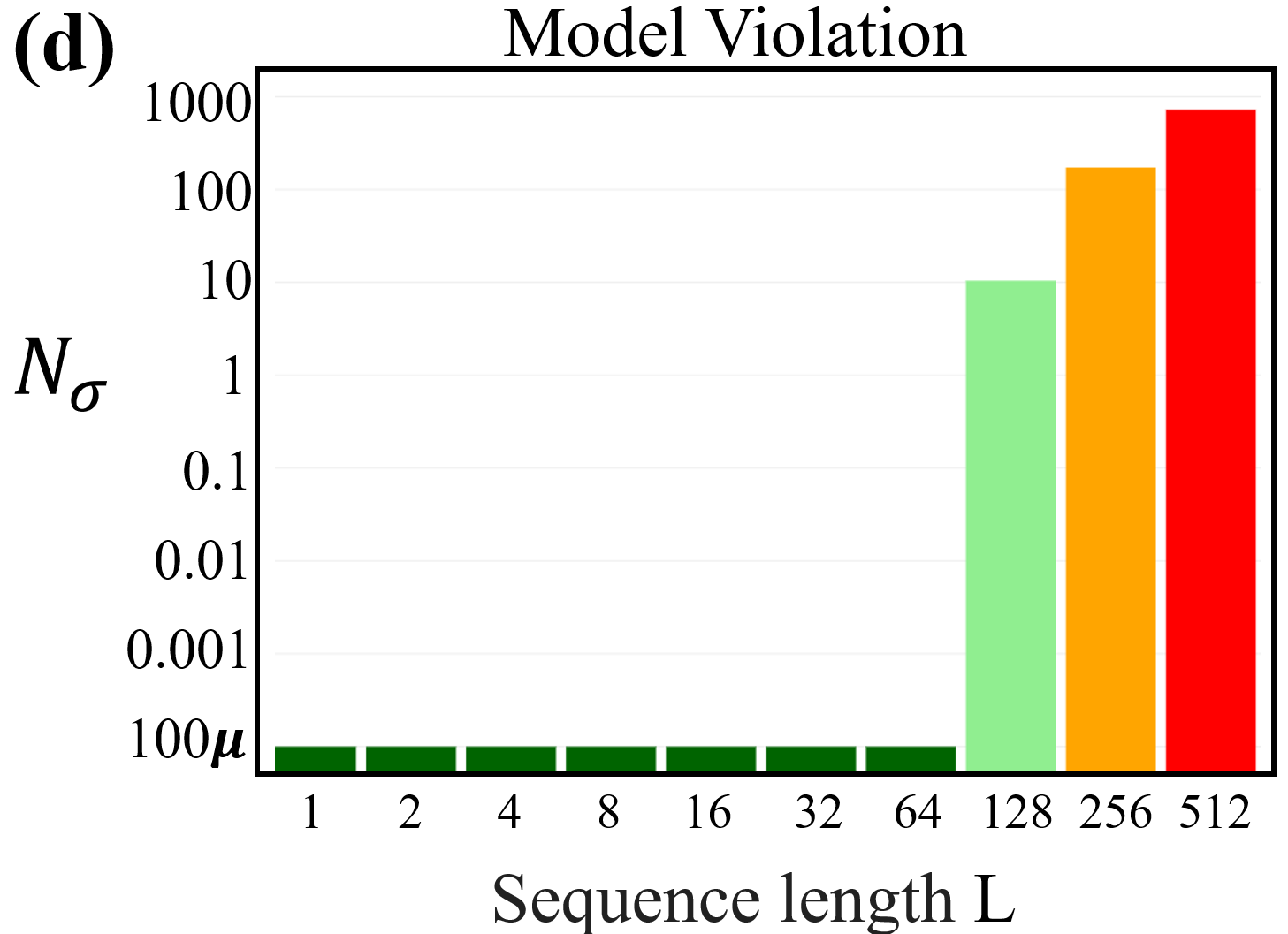}
\end{subfigure}
\caption{\label{fig3} 
Error magnitudes (coefficients) and contributed average gate infidelity of the error generators along the axes of the Bloch sphere (X, Y, or Z) from the GST results using the same parameters as in Sec.\ref{sec:level4.1}. In the GST analysis, we assume the ideal gate operation followed by these error generators with their error magnitudes. The left columns of plots (a), (b), and (c) represent the error magnitudes of the error generators for the identity gate, $X/2$ gate, and $Y/2$ gate, respectively. The coherent Hamiltonian error generators are represented in blue, and the incoherent stochastic error generators are in ginger. The right column shows the average gate infidelity contributed by each error generator after the ideal gate operation. As Eq. (B5) indicates, stochastic errors generally contribute more than Hamiltonian errors because of the quadratic coefficient contribution of the Hamiltonian generators. The average gate fidelities of the identity gate, $X/2$ gate, and $Y/2$ gate are, respectively, $F_{I} = 99.941\%$, $F_{X/2} = 99.965\%$, $F_{Y/2} = 99.964\%$. (d) The model deviation metric $N_{\sigma}$ increases rapidly as germ sequences repeat. A larger $N_{\sigma}$ indicates greater deviation from the Markovian noise model assumed by the GST.}
\end{figure}

In a GST result, the error is attributed to the combination of various quantum channels, while the SPAM errors are excluded. Detailed error channels can be obtained by analyzing the error generators of each gate. We applied our quantum noise model, using the same parameters as described in \ref{sec:level4.1}, to examine its effect on the GST protocol. Since multiqubit GST analysis is a little bit complex and infidelity is dominated by Eq.~(\ref{entanglement_infidelity}), we only take a single-qubit gate set $\mathcal{G} = \{I/2, X/2, Y/2, Z/2\}$ and calculate only Hamiltonian error generators ($H_{P}[\rho]$) and stochastic error generators ($S_P[\rho]$), i.e., H+S mode analysis. Moreover, Z rotations can be implemented virtually in software by applying additional phase manipulation to X and Y rotations; they typically have much higher fidelity \cite{Mdzik2022, arxiv.2410.15590}. Therefore, we treat the Z/2 gate as ideal \cite{PhysRevA.96.022330}. In our GST result, we perform an analysis on each of the gates $\{I/2, X/2, Y/2\}$ operated at the resonance frequency and with the same operation time. The remaining three gates have different error generators, as shown in Fig.~\ref{fig3}. Although parts of $H_P[\rho]$ are sometimes large due to coherent noise such as quasi-static detuning or pulse deformation, the dominant contribution of entanglement infidelity comes from $S_P[\rho]$ due to Eq.~(\ref{entanglement_infidelity}). In this equation, the coherent part,$H_P[\rho]$, will be significantly reduced by its quadratic contribution to the process (entanglement) infidelity. In contrast, the contribution of stochastic error generators, $S_P[\rho]$, is linear and therefore dominates most of the infidelity. In addition, we also observe that the average gate fidelity in Fig.~\ref{fig3} shows high consistency with the gate fidelity from the RB experiment \cite{Yoneda2017}, which we refer to. This is because we have another quantum channel to mitigate the error contribution from the Hamiltonian error generators, thereby avoiding the introduction of excessive noise strength to the coherent noise. 

In the error generators of the identity gate, $I/2$, we readily observe the sources of the noise and the direction of classical and quantum noise. The error generators in the z direction at the gate $I/2$ can be attributed to our noise model. However, in the $X/2$ and $Y/2$ gates with pulses, the errors move almost half of the contributions to another direction. This suggests that fluctuations in the rotational gates at the resonance frequency may cause a tilted rotation, thus randomizing the rotation angle. The convolution result might contribute to a more complex effect on gate operation. This provides a possibility to manipulate the error generators using a pulse for further gate optimization. Since the ratio of error generators according to different taxonomies and directions is also consistent with the GST result of other real devices \cite{Huang2024, arxiv.2307.12452, PhysRevApplied.17.024068}, our model can explain the mechanism behind the GST result of the device and get closer to the real device.

On the other hand, the non-Markovian impact becomes significant as we observe the increasing violation of the GST model with the gate sequence in Fig.~\ref{fig3} (d). The metric $N_{\sigma}$ of GST model violation, defined in Appendix \ref{appendix E}, indicates the deviation from the applied Markovian error generator. GST analysis may not always be as accurate and helpful as we thought. Because PyGSTi calculates errors based on the maximum likelihood algorithm and the Markovian assumption, its analysis results generally do not show model violation for short gate sequences. However, standard GST analysis presented in non-Markovian noise for a long sequence will fail. By observing the growing trend in the violation of the GST model, the non-Markovian effect becomes stronger as the sequence length increases. This means that the magnitudes of the error generators in the latter part become stronger than the former parts of the gate train; therefore, the increasing magnitude of the error may be an obstacle against the GST analysis of non-Markovian noise \cite{Wang2024}. When applying GST to an operation, we must use longer sequences to evaluate its performance, as the non-Markovian effect can mislead the observation at short sequence lengths. A promising method for investigating longer quantum channels should also consider non-Markovian error generators, not only for Si quantum dot qubits but also for superconducting qubits \cite{li2024,vandenBerg2023, White2020}. Developing a tomography that is compatible with non-Markovian noise will be a crucial step towards achieving higher gate fidelity.

\subsection{\label{sec:level4.3}Gate optimization}

In our quantum noise model, we attribute a portion of the error to the incoherent error that occurs when operating the quantum gate. Typically, incoherent Markovian errors can only be reduced by reducing the gate operation time, and some slow, incoherent, non-Markovian errors can be suppressed by pulse reshaping optimization \cite{PhysRevA.85.032321, Yang2019-yk}. To achieve higher gate fidelity and approach fault-tolerant quantum computing, efficient qubit calibration and gate optimization in the presence of noise are crucial steps.  Here, we demonstrate that our stochastic error generator, which is considered an incoherent, non-Markovian error with a $1/f$ noise spectrum, can be further suppressed through gate optimization. Moreover, the optimized reshaped pulse is demonstrated to be more robust against coherent noise than the regular Gaussian pulse through a filter function analysis in a CPMG protocol, highlighting the significant effectiveness of the optimized pulse.

Following Krotov's algorithm, we set the coherent noise strength $\delta(t)$ to zero to create a simplified scenario to test whether the effect of the incoherent error can be suppressed. We define gate infidelity $F_{\text{inf}} = 1-F$ with process fidelity, $F = Tr(\hat{Q}^{\dagger}G)/N$ where $N$ is the dimension of the propagator matrix $G$. We define a normal Gaussian pulse 
\begin{align}
\label{ini_guess}
    \Omega_{\text{Gau}}(t) = \Omega_0e^{-\frac{(t-0.5T)^2}{0.125T^2}},
\end{align} 
where the amplitude $\Omega_0$ is defined by an ideal $Y/2$ gate (without noise) during gate time $T$. Then, using this ideal normal Gaussian pulse for a gate time $T=15$ ns, we obtain a gate infidelity $F_{\text{inf}} = 0.017\%$ for a $Y/2$ gate in the incoherent noise environment.

\begin{figure}
\begin{subfigure}
  \centering
  \includegraphics[width=.48\linewidth]{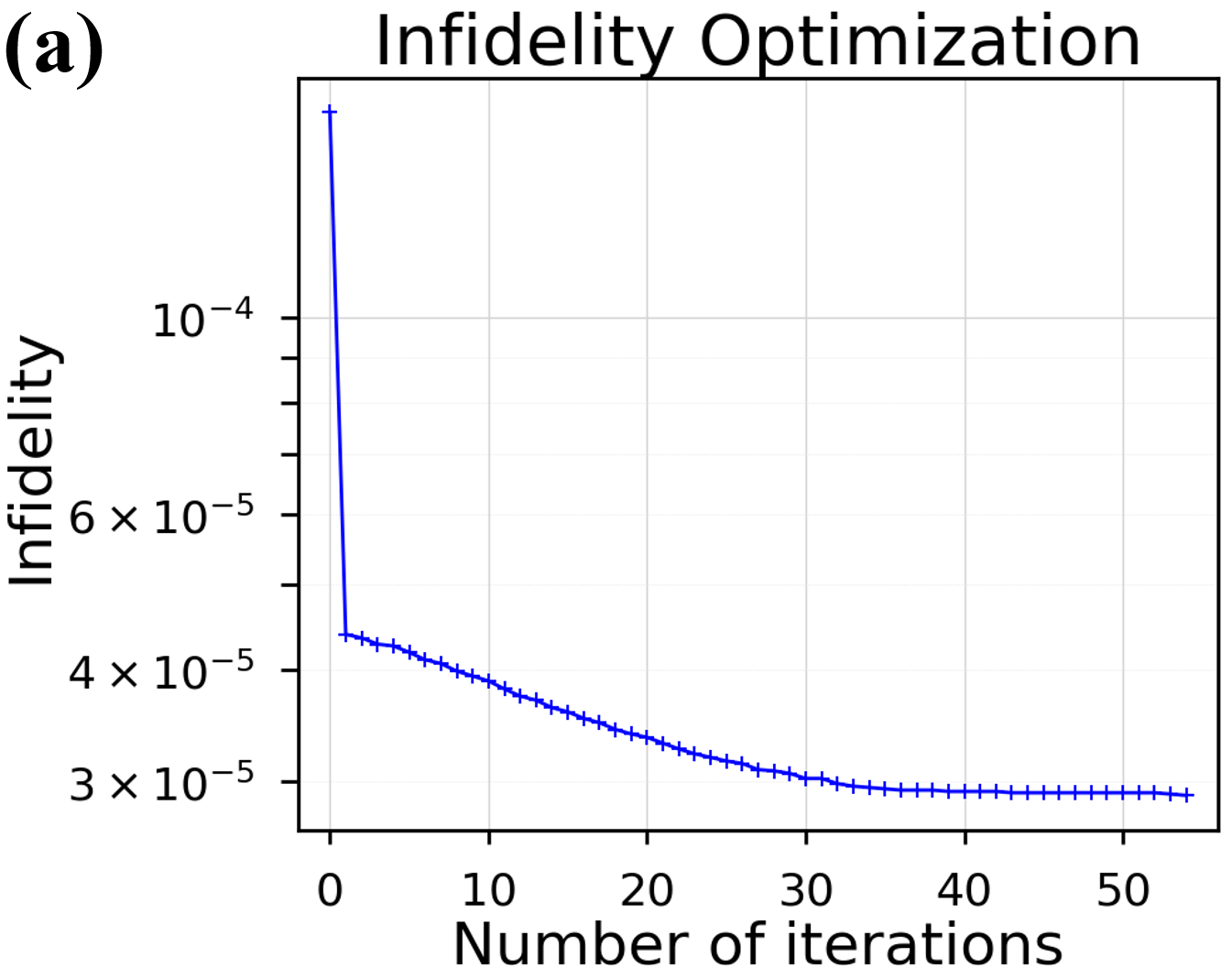}
\end{subfigure}%
\begin{subfigure}
  \centering
  \includegraphics[width=.48\linewidth]{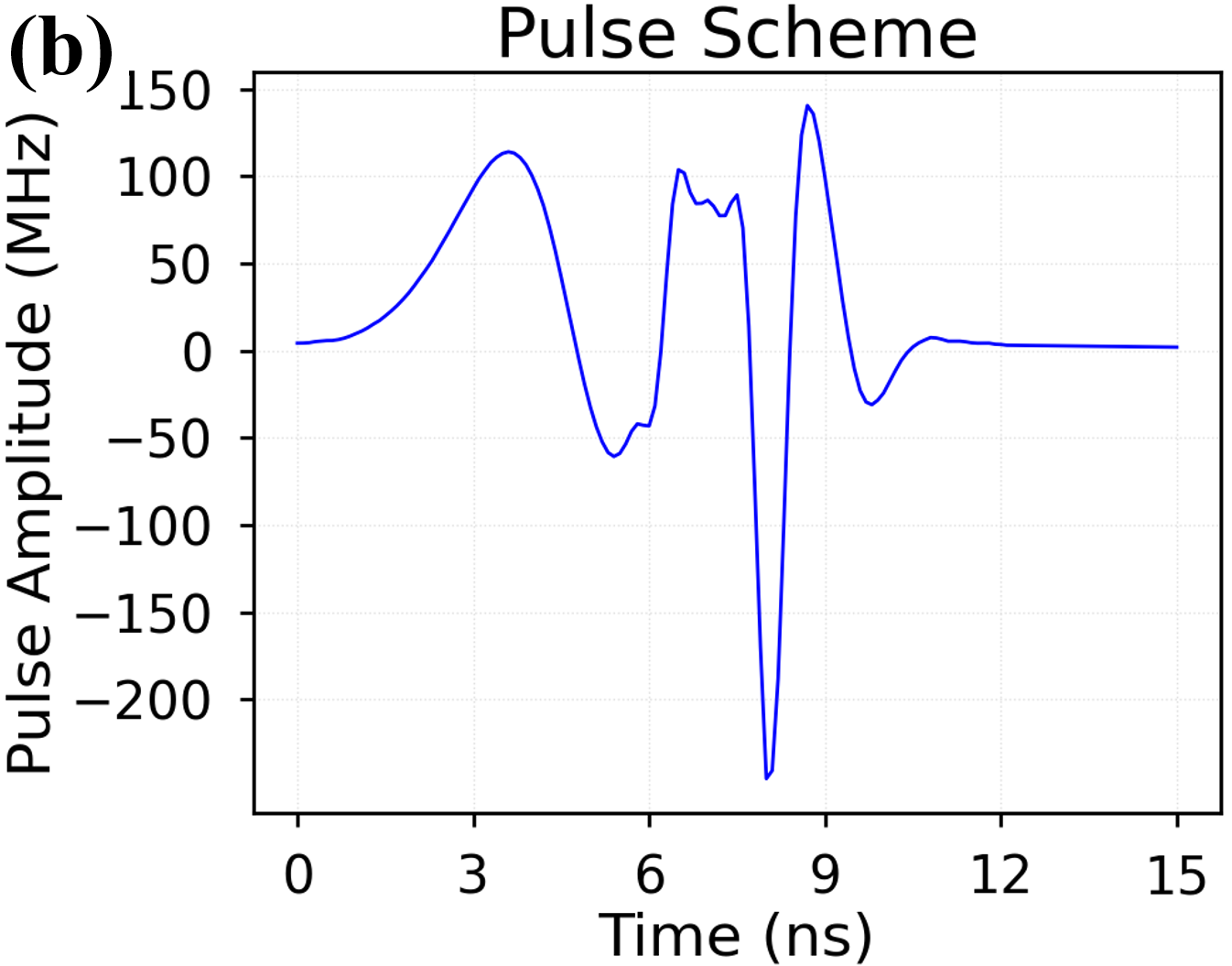}
\end{subfigure}
\begin{subfigure}
  \centering
  \includegraphics[width=.48\linewidth]{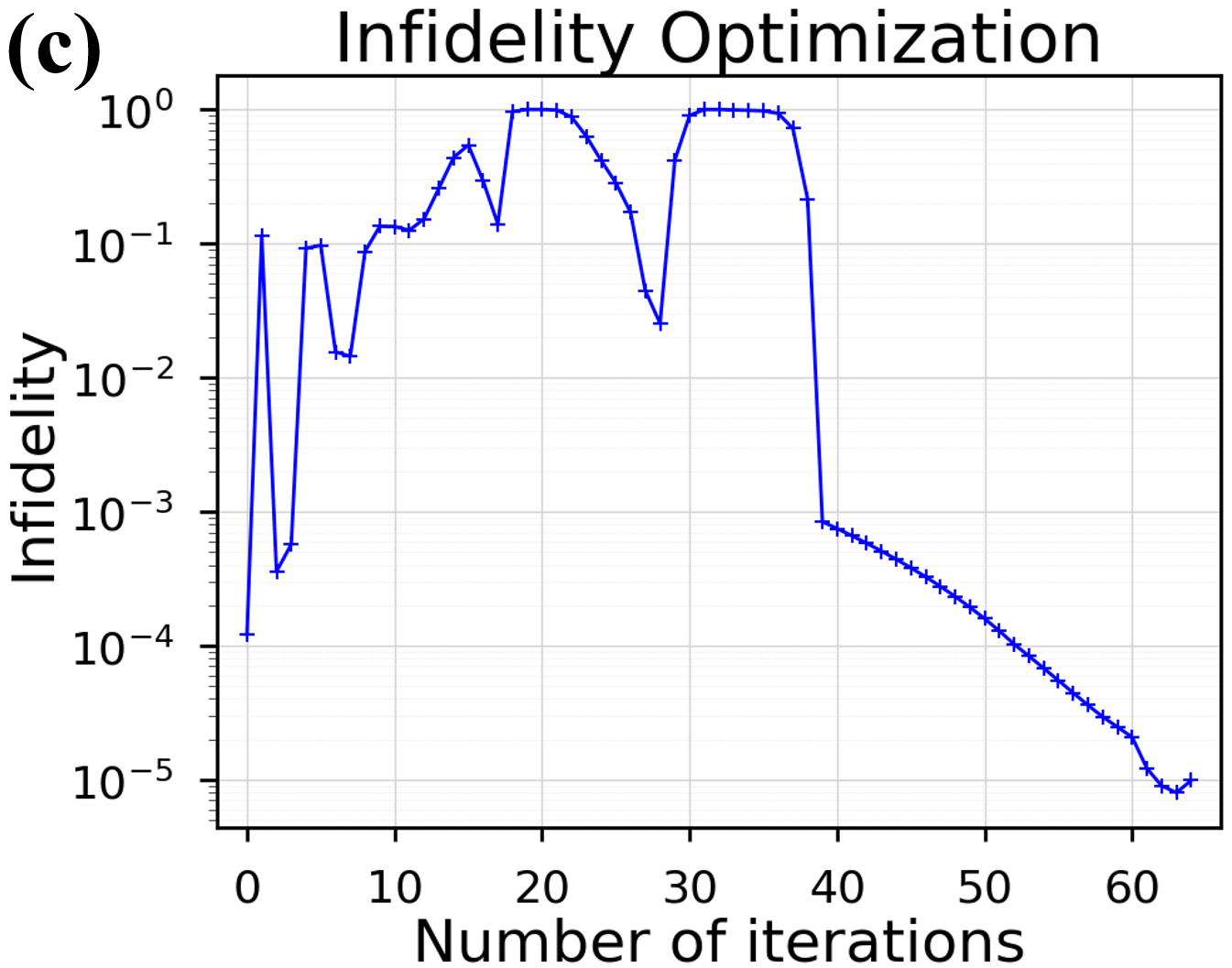}
\end{subfigure}
\begin{subfigure}
  \centering
  \includegraphics[width=.48\linewidth]{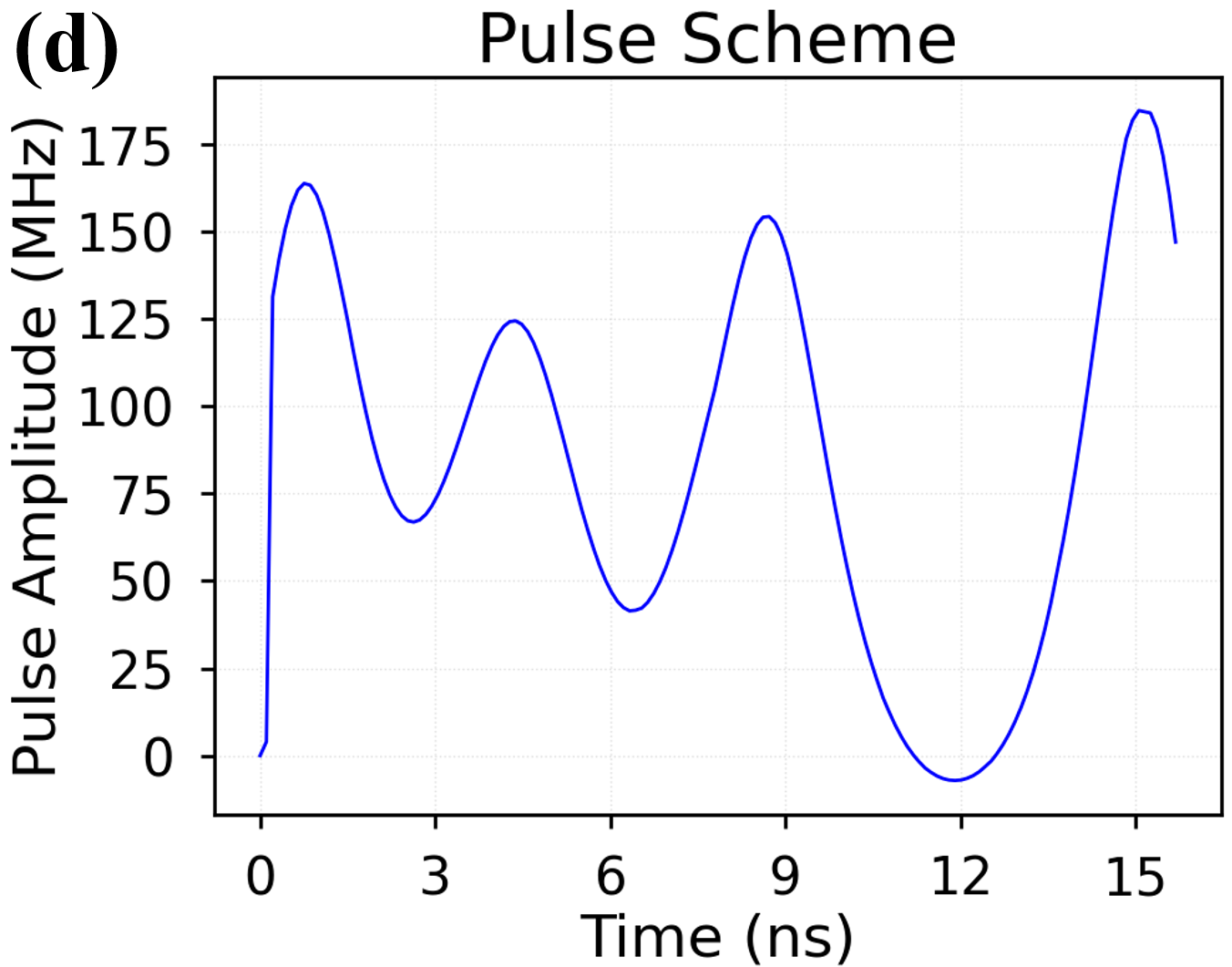}
\end{subfigure}
\caption{\label{fig4}Pulse optimization results of the $Y/2$ gate for two cases: (a) and (b) for the case where the local minimum is near the original start point, and (c) and (d) for the case showing crossing over the original local minimum to find a new local minimum via adjusting the step size based on the values of infidelity. In (a), the gate infidelity that is calculated from process fidelity decreases monotonically following the iteration of optimization, and the infidelity $F_{\text{inf}}$ decreases from 0.017\% to 0.0028\%, depicted on a logarithmic scale. Illustration (b) is the fine-tuning pulse shape corresponding to the best fidelity $F=99.9972\%$. In (c), the figure is depicted on a logarithmic scale because the optimization was set to overcome the barrier near the initial local minimum, so the infidelity ranges over a large order of magnitude. The lowest reachable infidelity is $ F_{\text{inf}} = 0.00079\%$. (d) The pulse reaching a new local minimum at fidelity $F = 99.99921\%$ has a quite different shape from the original Gaussian function.}
\end{figure}

Next, we demonstrate our optimized result. At the end of Sec.~\ref {sec:level3.2.2}, we mentioned that $1/\lambda(t)$ in the objective function Eq.~(\ref{obj_func}) can also be considered as the step size in Eq.~(\ref{iteration}). We select the normal Gaussian pulse, $\Omega_{\text{Gau}}(t)$, as the initial pulse and a duplicate multiplied by 0.02 as the step size, $[1/\lambda(t)]=0.02\Omega_{\text{Gau}}(t)$. After undergoing the optimization process of Krotov's algorithm, the optimization result and the reshaped pulse are shown in Figs.~\ref{fig4} (a) and \ref{fig4} (b). Since the initial fidelity is high, the local minimum is close to the starting point. The infidelity therefore decreases rapidly, and then the small step size sustains the iteration, monotonically roaming near the local minimum. Until the preset limit of stability is reached, the iteration reaches the lowest infidelity $F_{\text{inf}} = 0.0028\%$.

To improve optimization speed and achieve better gate fidelity, we changed some settings in our optimization process. In the first trial, the optimization roamed around the beginning local minimum. So, we tried to escape this local minimum and cross the surrounding barrier to find a better minimum. Increasing or decreasing the amplitude of the step size in a Gaussian waveform did not yield a better solution. We observe in Fig.~\ref{fig4} (b) that there are multiple oscillations in the pulse length, although we did not add any sinusoidal function in the step function, $1/\lambda(t)$, suggesting that the optimization was trying to utilize an oscillating pulse to suppress errors. This inspired us to use a sinusoidal function for further optimization. Additionally, to balance exploration and convergence—akin to learning rate scheduling in machine learning—we adopt a dynamic modulation strategy for the step size to incorporate a progressive approach across different intervals of infidelity. We thus change the step size to a sinusoidal function $[1/\lambda(t)] = M \Omega_0 [\sin(\omega t)+1]$, where $\Omega_0$ is the amplitude from the initially guessed Gaussian pulse Eq.~(\ref{ini_guess}), $\omega=1/T$ is the inverse of the gate time, and $M$ is a dynamic number adjusted according to the different infidelity intervals below. For high infidelity intervals, such as $F_{\text{inf}}>10^{-2}$, we use a Gaussian pulse with a scaled amplitude of 0.1 to pass through the local minima. For smaller infidelities, we use a smaller sinusoidal function and dynamically adjust its amplitude based on the infidelity in each interval to vary the step size and accelerate the iteration. Specifically, we use the following step size strategy for different infidelity intervals:  
\begin{alignat}{2}
&\frac{1}{\lambda(t)} = 0.1\Omega_{\text{Gau}}(t),
\hspace{40pt} F_{\text{inf}}>10^{-2}\,\,\text{or}\,\,F_{\text{inf},n}>F_{\text{inf},n-1}\\
&\frac{1}{\lambda(t)} = 0.01 \Omega_0 \frac{\sin(\omega t)+1}{[1/(F_{\text{inf}} \cdot 10^3)]+1},  \hspace{17pt}10^{-3}<F_{\text{inf}}<10^{-2}\\
&\frac{1}{\lambda(t)} = 0.002 \Omega_0 \frac{\sin(\omega t)+1}{[1/(F_{\text{inf}} \cdot 10^4)]+1},  \hspace{12pt}10^{-4}<F_{\text{inf}}<10^{-3}\\
&\frac{1}{\lambda(t)} = 0.001\Omega_0\frac{\sin(\omega t)+1}{[2/(F_{\text{inf}} \cdot 10^5)]+1},  \hspace{12pt}10^{-5}<F_{\text{inf}}<10^{-4}\\
&\frac{1}{\lambda(t)} = 0.0005\Omega_0\frac{\sin(\omega t)+1}{[2/(F_{\text{inf}} \cdot 10^6)]+1},  \hspace{40pt}F_{\text{inf}}<10^{-5}.
\end{alignat}
In each infidelity interval, we add a factor to the denominator to adjust the step size based on the current infidelity. When approaching low infidelity, we want the step size to be smaller to avoid missing the minimum infidelity. But a small step size requires more iterations and thus time to converge. Therefore, the simple design of the denominator in each infidelity interval enables a larger step size when infidelity is greater and a smaller step size when it is lower. Figures~\ref{fig4} (c) and \ref{fig4} (d) show the results of our strategy to find a new local minimum. In Fig.~\ref{fig4} (c), the variation of infidelity in about half of the total iterations depicts the process to overcome the barrier near the initial local minimum with a Gaussian step size. For the remaining half, Krotov's algorithm, accompanied by the new strategy, can decrease infidelity stably and continuously as it approaches the low-infidelity interval. The optimized pulse shown in Fig.~\ref{fig4} (d) has a lower infidelity $F_{\text{inf}} = 0.00079\%$ and a quite different waveform from that in Fig.~\ref{fig4} (b), indicating a successful escape from the original local minimum. The oscillations in the pulse may be attributed to the fact that the operator $\mathcal{K}(t)$ of Eq.~(\ref{dissipation_kernel}) can be manipulated to reduce the error accumulation rate, similar to dynamical decoupling.

\begin{figure}[!ht]
\centering
\includegraphics[width=.9\linewidth]{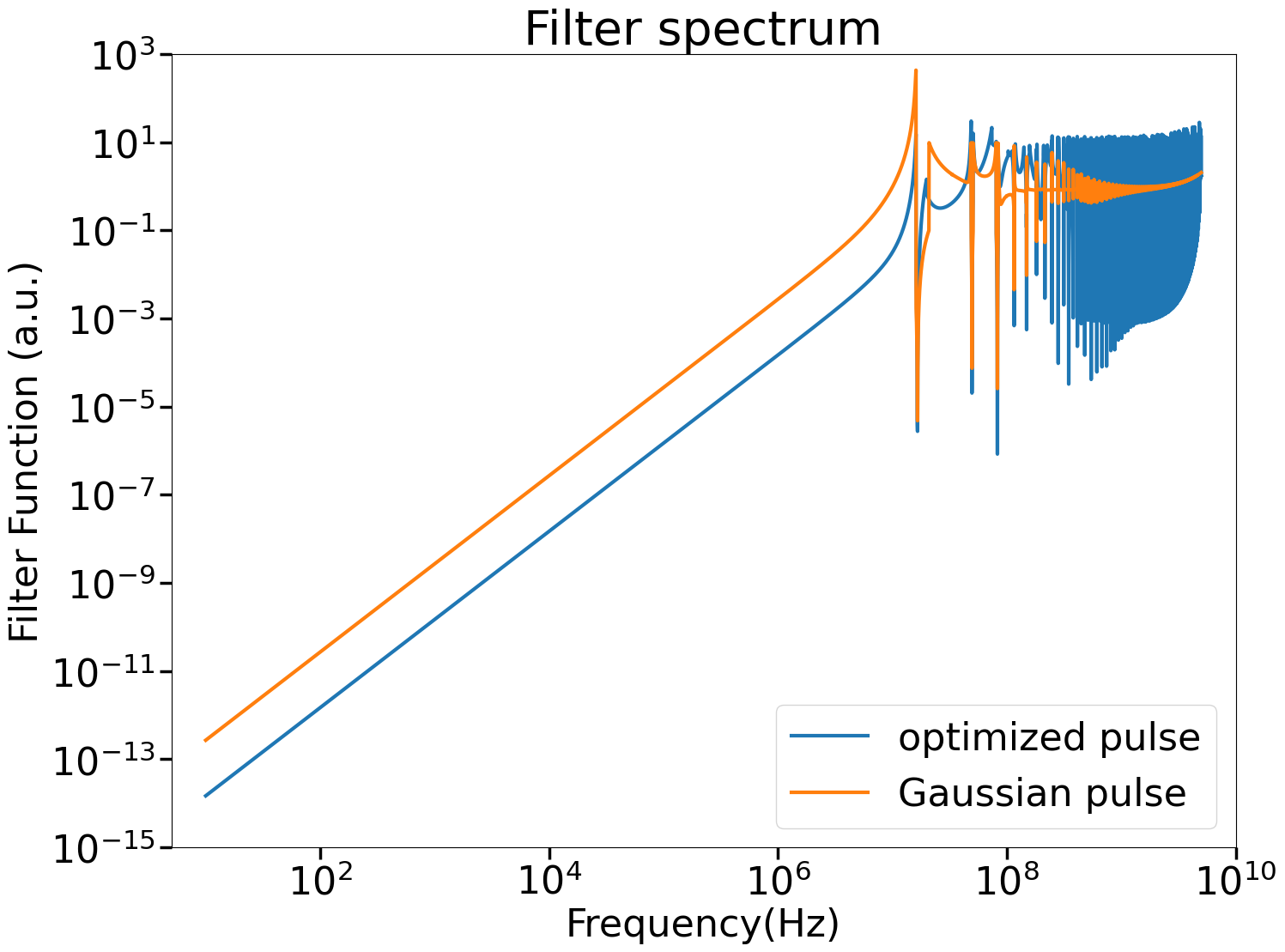}
\caption{\label{fig5} The normalized filter spectrum converted from the control matrix $[\mathbf{R}(t)]_{ij}$ presents how much charge noise would be removed in the operation process. The orange line represents the reference for our comparison, using a Gaussian pulse as the initial pulse in the optimization. The blue line is the result of using the optimized pulse depicted in Fig.~\ref{fig4} (d).}
\end{figure}

In the previous optimization, we disabled the classical noise to facilitate efficient optimization. To determine the ability of the optimized pulse to mitigate coherent errors, we demonstrate the difference in the filter functions obtained in the CPMG protocol with Y gates using both the regular Gaussian pulse and the optimized pulse in the presence of classical coherent noise. The average gate fidelity that suffers from the classical dephasing noise is approximated in \cite{Green2013}

\begin{align}
\label{fidelity}
F_{\text{avg}}\approx \frac{1}{2}(1+e^{-\chi(\tau)}), 
\end{align} 
and the decay rate of gate fidelity takes the form:
\begin{align}
\label{decay_rate}
    \chi(\tau) = \frac{1}{\pi}\int^{\infty}_0\frac{1}{\omega^2} S_z(\omega) F_z(\omega) d\omega, 
\end{align}
where $S_z(\omega)$ is the noise spectrum and 
\begin{equation}
F_z(\omega)=\sum_j |R_{zj}(\omega)|^2
\end{equation} 
is the filter function extracted from the control matrix 
\begin{equation}
    R_{ij}(t) = Tr(U^{\dagger} \sigma_i U \sigma_j)/2, \quad i,j = \{x,y,z\},
\end{equation}
i.e., the gate matrix in the Pauli transfer formalism, and $\tau$ represents the time elapsed during the process. If the filter function is reduced by a special pulse scheme, the average gate fidelity will decay more slowly and therefore be sustained at a higher level than before. The filter function can be used for comparing different dynamical decoupling sequences without introducing real noise or for comparing different pulse waveforms in the same noise environment and the same dynamical decoupling sequence. In Fig.~\ref{fig5}, within a CPMG protocol, we test the difference in the filter function obtained in the presence of $1/f$ charge noise between a normal Gaussian pulse and the optimized pulse that yields the best fidelity, $F = 99.9992\%$, for the Y/2 gate. In Fig.~\ref{fig5}, there is a significant reduction in the filter function spectrum in the low-frequency regime. Furthermore, the blue line representing the use of the optimized pulse for the $\pi$ (Y) gate in Fig.~\ref{fig5} moves down from the orange line representing the use of a Gaussian pulse in parallel before the frequency reaches $10^7$ (Hz), which also corresponds to the case where the maximum number of $\pi$ (Y) gates is implemented in the CPMG sequence.
The result indicates that the optimized pulse has a better ability to reduce the effect of classical dephasing noise in any of the stated frequency regions. In other words, even though our control pulse is optimized for quantum (incoherent) noise of the $1/f$ noise spectrum, it is more robust against the classical dephasing noise of a similar $1/f$ noise spectrum than a regular Gaussian pulse.

\section{\label{sec:level5} Conclusion}

We have investigated the $1/f$ noise problem that appears in the Si quantum dots and affects the gate operations. As we strive for fault-tolerant quantum computing in the future, it is essential to incorporate optimized control and measurement to achieve high-fidelity quantum gates. Addressing the challenge of $1/f$ charge noise is crucial in our efforts to engineer the quantum gate. Using the GST method has provided us with comprehensive information on the gate matrix, allowing us to identify the main sources of infidelity. The introduced stochastic quantum spin-boson model, which aligns with the classical coherent noise model typically modeled only through the system Hamiltonian, along with the quantum master equation, effectively captures the decoherence behavior of the qubit. Our focus on accounting for environmental influences is pivotal in mitigating non-Markovian noise, such as $1/f$ noise. Finally, we used the Krotov method to optimize the single-qubit gate operation, resulting in a reduction in gate infidelity of approximately an order of magnitude. Furthermore, in the engineering of two-qubit gates, such as CZ gates, it is crucial to identify error mechanisms and develop an effective noise model prior to pulse optimization. Integrating our model with pulse engineering offers a viable path to enhance average gate fidelity while concurrently reducing incoherence errors. Ultimately, pinpointing the relevant noise sources, understanding and characterizing their spectra, and suppressing their effects on quantum gate operations are key steps toward achieving fault-tolerant quantum computers.

\begin{acknowledgments}
W.-E.C would like to acknowledge support and encouragement from I-Chun Hsu, Ssu-Chih Lin, and Chien-Chang Chen for their suggestions in the methodology and the optimization algorithm.
H.-S.\ Goan acknowledges support from the National Science and Technology Council (NSTC), Taiwan, under Grants No.\ NSTC 113-2112-M-002-022-MY3, No.\ NSTC 113-2119M-002-021, No.\ NSTC 114-2119-M-002-018, No.\ NSTC 114-2119-M-002-017-MY3, and NSTC 115-2119-M-002-005,  the support of Taiwan Semiconductor Research Institute (TSRI) through the Joint Developed Project (JDP), and from the Physics Division, National Center for Theoretical Sciences (NCTS), Taiwan. H.-S.\ Goan also acknowledges support from the National Taiwan University (NTU) under Grants No.\ NTU-CC-115L8937, No.\ NTU-CC-115L893704 and No.\ NTU-CC-115L8512, as well as support from the Center for Advanced Computing and Imaging in Biomedicine (NTU-115L900702) through the Featured Areas Research Center Program within the framework of the Higher Education Sprout Project by the Ministry of Education (MOE), Taiwan.

\end{acknowledgments}
\appendix
\section{\label{appendix A} EDSR and interaction Hamiltonian}

\begin{figure}[h!]
\includegraphics[ width = 0.45\textwidth]{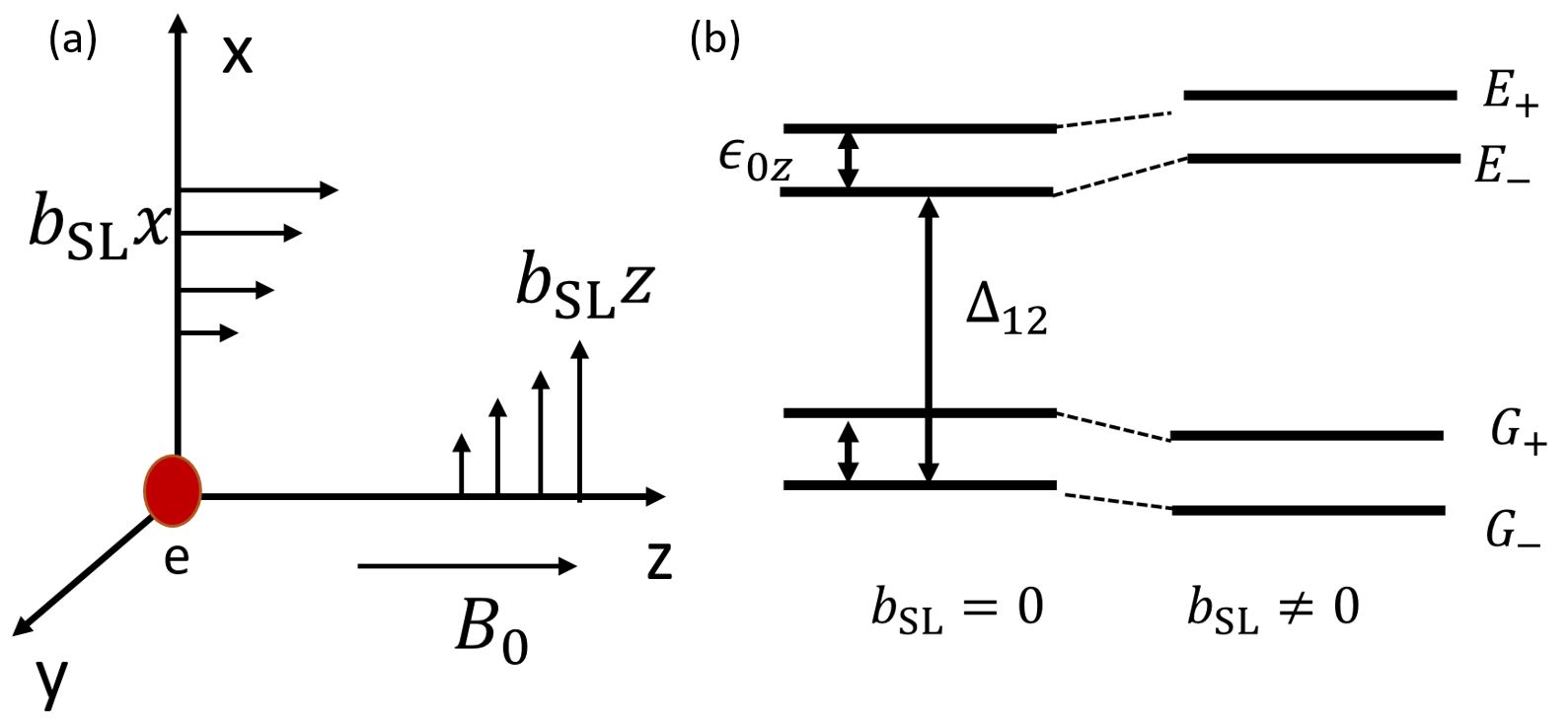}
\caption{\label{fig1} (a) The schematic of EDSR of a single electron. A static magnetic field $\vec{B}_0$ is applied along the z direction, and a micromagnet also contributes a magnetic field gradient $b_{\text{SL}}$. By applying AC voltage, the single electron oscillates in the x-z plane.  (b) The left side is the original energy level when there is only a parabolic quantum dot potential. The right side is the new energy level calculated by perturbation theory after adding an inhomogeneous field. Here, $G$ denotes the orbital ground state and $E$ represents the orbital excited state. The term $\Delta_{12}$ refers to the orbital energy, typically set at 1 meV when the dot's dimensions are restricted to approximately 50 nm.}
\end{figure} 

We describe here the basic operation principle of silicon-based quantum-dot electron spin qubits and outline the procedure to obtain the effective spin-phonon coupling model in Eq.~(\ref{total_H}) in the main text. A single electron is placed within the x-z plane of a quantum dot in a static magnetic field $\vec{B}_0$ applied along the z direction, as illustrated in Fig.~\ref{fig1}. The electron dipole spin resonance control scheme (EDSR)  uses a micromagnet to generate an inhomogeneous magnetic field gradient, resulting in a total magnetic field $ \vec{B} = b_{\text{SL}}z \hat{x} + (B_0 + b_{\text{SL}}x) \hat{z}$, where $b_{\text{SL}}$ denotes the field gradient along the direction z (x), parallel to the x (z) axis. This approach can be extended to perform controlled rotations for two-qubit gate operations \cite{doi:10.1126/science.aao5965, Noiri2022-xw}. The orbital energy of the electron within the quantum dot is constrained by the electric potential of the barrier gate located above the dot. The inhomogeneous magnetic field generated by the micromagnet, as shown in Fig.[\ref{fig1}], couples the spin degrees of freedom to the orbital  \cite{PhysRevLett.96.047202, PioroLadrire2008, PioroLadrire2007}.
The Hamiltonian describing an electron in a parabolic quantum dot potential in the above setup is as follows:
\begin{align}
\begin{split}
    \hat{H}_{\text{sys}} &= \hat{H}_{\text{0o}} + \hat{H}_{\text{0s}} \\
    &= \frac{\hat{p}^2}{2m} + \frac{1}{2} m \omega_0^2 \hat{z}^2  -g\mu_{\text{B}} B_0 \hat{S}_z  - g\mu_{\text{B}} b_{\text{SL}}z \hat{S}_x. 
\end{split}
\end{align}
Here, the first two terms represent the orbital energy, including the kinetic and potential energy within the dot. The latter part is the Zeeman energy in an inhomogeneous field. $g$ is the effective $g$-factor of an electron in silicon, $\mu_{\text{B}}$  is the Bohr magneton, and $\hat{\vec{S}}=\vec{\sigma}/2$ is the Pauli spin operator. After applying perturbation theory, the Hamiltonian in the subspace of the orbital ground state $\ket{G_{\sigma}}$ is written as \cite{PhysRevLett.96.047202}
\begin{align}
\label{micro_magnent_H}
    &\hat{H}_s = \frac{1}{2} \epsilon_{0z}\left(1- \frac{M_{12}^2}{2(\Delta_{21}^2-\epsilon_{0z}^2)}\right) \sigma_z,
\end{align}
where $\epsilon_{0z}$ comes from the Zeeman effect, and the remaining parameters $M_{12}, \Delta_{12}$ are related to the magnetic gradient orbital eigenstate and eigenenergy. In our model, we do not consider additional time-dependent magnetic-field fluctuations. The correlation time of the hyperfine magnetic noise arising from the nuclear spins of the $^{29}\text{Si}$ isotope is sufficiently long to allow a series of gate operations to be completed with feedback control, as demonstrated in \cite{Yang2019-yk}. Therefore, we focus on the shorter timescale relevant to the charge noise and treat the effect of the hyperfine interaction as effectively constant in our analysis of quantum gate operations \cite{Yang2019-yk, PRXQuantum.5.010306}.

We describe a pathway to illustrate how charge noise can influence electron spin in the EDSR control scheme. We model an atom, near the interface, separated by an ion and an additional electron. The distribution of the charge separations of atoms can be viewed as a polarization density, allowing us to write down an interaction from the Coulomb potential in the second-quantization formalism using field operators. The second quantized operator in the multipole expansion for a fluctuation in displacement of the polarization density $\hat{u}$ composed of the phonon annihilation and creation operator: $\hat{u}_j = \sum_k u_k(\hat{a}_k+\hat{a}_{-k}^{\dagger})$, where $u_k$ is the displacement of each phonon mode. The Coulomb interaction-mediated electron-phonon coupling can be described with the Jellium model. 

Since spin couples to the position due to the inhomogeneous magnetic field gradient in EDSR, we adopt a semiclassical approximation, as used in \cite{Yoneda2017, Yoneda2023}, to calculate the energy change. Given a fluctuating electric field $\delta E_{\text{rms}}$ at the interface, the electron in the quantum dot will tilt slightly. In the presence of an inhomogeneous magnetic field, the Hamiltonian that couples spin and orbital degrees of freedom contributes to charge noise. The potential in the Hamiltonian can be approximated as:
\begin{align}
    \hat{V}_{\text{orbital}} =  e\delta E_{\text{rms}} r + \frac{1}{2}m\omega_0^2 r^2 =  \frac{1}{2}m\omega_0^2(r+\delta r)^2,
\end{align}
where $r$ is the displaced distance.
In the first-order expansion, the tilted distance is given by $\delta r = \frac{e}{m\omega_0^2}\delta E_{\text{rms}}$. In addition, the inhomogeneous magnetic field $ \vec{B} = b_{\text{SL}}z \hat{x} + (B_0 + b_{\text{SL}}x) \hat{z}$ couples the spin-orbital energy. If we let this tilting deviation be along the $x$ direction, the quantum-dot displacement will shift the spin qubit frequency via a longitudinal magnetic field by
\begin{align}
\label{semi_classical_approx}
    \hbar \delta_{\text{c}} = g\mu_{\text{B}}b_{\text{SL}} \delta r,
\end{align}
and this detuning will add to the $\epsilon_{0z}$ of Eq.~(\ref{micro_magnent_H}) as the first term in the multipole expansion. The fluctuating electric field will correspond to an equivalent voltage noise. In this manner, and with the classically observed voltage noise, the charge-noise detuning can be estimated to lie in the range $\delta_{\text{c}} \approx 10 \sim 100$ kHz, depending on the slanting magnetic gradient. After finishing the first term, which corresponds to the classical voltage noise energy, we then add the second term induced by the dipole field following the previous polarization density fluctuation assumption, and use a mean-field approximation:
\begin{align}
\begin{split}
    H_{\text{int}} &= \hbar \delta_{\text{c}} \sigma_z + e \delta E_{\text{rms}} \hat{u} \sigma_z\\
    &= \hbar \delta_{\text{c}} \sigma_z + \sum_k e \delta E_{\text{rms}} {u_k}  \sigma_z (\hat{a}_k+\hat{a}^{\dagger}_{-k})\\
    &=\hbar \delta_{\text{c}} \sigma_z + \sum_k \hbar g_k \sigma_z (\hat{a}_k+\hat{a}^{\dagger}_{-k}).
\end{split}
\end{align}

By considering the classical noise coming from time-dependent fluctuating charge or voltage noise by replacing $\delta_{\text{c}}\Rightarrow \delta_{\text{c}}\nu(t)$ and adding the phonon energy, the effective noise interaction Hamiltonian can be written as
\begin{align}
    H_{\rm int} = \hbar \delta_{\text{c}}\nu(t) \sigma_z + \sum_k \hbar \omega_k \hat{a}^{\dagger}_k\hat{a}_k +\sum_k \hbar g_k \sigma_z (\hat{a}_k+\hat{a}^{\dagger}_{-k}),
\end{align}
which is the model Hamiltonian that describes the classical charge noise and spin-phonon coupling Hamiltonian in Eq.~(\ref{total_H}) of the main text.

\section{\label{appendix B} Error generator in GST}

 Here, we briefly introduce some of the basics and benchmarking properties extracted from the GST analysis. In the Markovian approximation, every unitary operation can be written as an ideal unitary gate followed by an error quantum channel \cite{PRXQuantum.3.020335, Nielsen2021}. Then a normal quantum channel takes the form of 
\begin{align}
\label{quantum channel}
    \mathcal{G}(\rho) = \mathcal{E} \circ U[\rho],
\end{align}
where $U[\cdot] = U \cdot U^{\dagger}$ is the target quantum channel and $\mathcal{E}$ is the error channel. Similarly, the state preparation and measurement(SPAM) can also be written as a perfect projection channel followed by a SPAM error channel.  Under the GST protocol, we can separately measure the gate and SPAM error channels (matrices). The error taxonomy for GST can be achieved by converting the noisy channels ($\mathcal{E}$) following each operation to their error generator($\mathbb{L}$) by assuming an infinitesimal error converging in the expansion:
\begin{align}
\label{error generator}
    \mathcal{E}[\rho] = e^{\mathbb{L}} \rho \approx \rho + \mathbb{L}[\rho].
\end{align}
There are four types of error generators, Hamiltonian ($H_{P}[\rho]$), stochastic ($S_{P}[\rho]$), Pauli-correlation ($C_{P,Q}[\rho]$), and active ($A_{P,Q}[\rho]$), where $P$ and $Q$ denote non-identity Pauli elements. Using PyGSTi \cite{Nielsen2021}, we project $\mathbb{L}$ into the full error space or the subspace of Hamiltonian and stochastic error generators and extract the coefficients of each elementary error generator as follows:
\begin{align}
    &\mathbb{L}[\rho] = \sum_P h_P H_P[\rho] + \sum_P s_P S_P[\rho] \\
    &+ \sum_{P,Q>P} c_{P,Q} C_{P,Q}[\rho] + \sum_{P,Q>Q} a_{P,Q} A_{P,Q}[\rho],  
\end{align}
where $h_P$, $s_P$, $c_{P,Q}$, and $a_{P,Q}$ are the coefficients determining the strengths of the corresponding elementary error generators, respectively.

Since the correlation error $C_{P,Q}$ and active error $A_{P,Q}$ contribute in quadratic form as $H_{P}$, depending on their commutating relation, when $[P,Q]=0$, $c_{P,Q} \neq 0$  and when $\{P,Q\}=0$, $a_{P,Q} \neq 0$, respectively. In spin qubits, the Hamiltonian errors and stochastic errors usually dominate in magnitude because $c_{P,Q}$, and $a_{P,Q}$ are extremely low. This is also the main reason why we focus only on Hamiltonian and stochastic errors in the GST. In this case, the entanglement infidelity (process infidelity) $ F_{\text{inf}} $ can be estimated based on these error coefficients, given by the sum over the extracted coefficients \cite{PRXQuantum.3.020335}:
\begin{align}
\label{entanglement_infidelity}
    F_{\text{inf}}=1-F_{\text{ent}} \approx \sum_P s_P + \sum _P h_P^2.
\end{align}
To make a connection between the interleaved RB and GST measurement, we can also calculate the average gate fidelity $F_{\text{avg}}$ as follows \cite{PhysRevA.60.1888}:
\begin{align}
    F_{\text{avg}} = \frac{d\cdot F_{\text{ent}}+1}{d+1},
\end{align}
where $d$ is the dimension of the Hilbert space (2 for a single-qubit system). Eq.~(\ref{entanglement_infidelity}) indicates that, to the gate infidelities, stochastic errors generally contribute more than Hamiltonian errors due to the quadratic coefficients of the Hamiltonian generators. Even in cases where the magnitudes of the Hamiltonian errors are significantly larger, this situation is commonly observed \cite{Huang2024}.

\section{\label{appendix C} Time-local master equation}
Here, we provide the derivation of the time-local master equation (\ref{master_eq}) and (\ref{dissipation_kernel}) as presented in the main text. Following the assumption of a factorized initial system-bath state $\rho_{tot}(0)=\rho(0)\otimes R_0$, the standard perturbative master equation under only the Born approximation in the interaction picture takes the form of 
\begin{equation}
\label{perturbation_expansion}
\begin{split}
\dot{\tilde{\rho}}(t)&=
-\frac{i}{\hbar}{\rm Tr}_B[\tilde{H}_I(t),\rho(0)\otimes R_0]\\
&-\frac{1}{\hbar^2}{\rm Tr}_B\int_0^t dt'[\tilde{H}_I(t),[\tilde{H}_I(t'),\tilde{\rho}(t')\otimes R_0]].
\end{split}
\end{equation}
Here $\tilde{\rho}(t)$ is the reduced density matrix of the system in the interaction picture, the initial thermal reservoir density operator at temperature $T$ is $R_0=\exp(-H_B/k_B T)/{\rm Tr}_B[\exp(-H_B/k_B T)]$, and the remaining Hamiltonian in the interaction picture is the interaction Hamiltonian of the spin-boson coupling,
\begin{equation}
\label{interaction_in_rotating}
\tilde{H}_{I}(t)
=\sum_k \hbar g_k\tilde{\sigma}_z(t)(\hat{a}_ke^{-i\omega_k t}+\hat{a}^{\dagger}_ke^{i\omega_k t}),
\end{equation}
where $\tilde{\sigma}_z(t)= U_S(t)\sigma_zU_S^\dagger(t)$ with the time-ordering system evolution operator $U_S(t)=T_+e^{-\frac{i}{\hbar}\int_0^t H_S(t')dt'}$. Substituting  Eq.~(\ref{interaction_in_rotating}) into Eq.~ (\ref{perturbation_expansion}) and then tracing out the reservoir degrees of freedom, we obtain the following result:
\begin{equation}
\label{master_eq_expansion_in_rotating}
\begin{split}
    \frac{d\tilde{\rho}}{dt} &= \int^t_0 dt' \{\left[\tilde{\sigma_z}(t')\tilde{\rho}(t') \tilde{\sigma_z}(t)-  \tilde{\sigma_z}(t) \tilde{\sigma_z}(t')\tilde{\rho}(t')\right] C(t-t') \\&+ \left[\tilde{\sigma_z}(t)\tilde{\rho}(t') \tilde{\sigma_z}(t')-  \tilde{\rho}(t')\tilde{\sigma_z}(t') \tilde{\sigma_z}(t)\right] C(t'-t)\},
\end{split}
\end{equation}
where the relation ${\rm Tr}_B[\tilde{H}_I(t)R_0]=0$ has been used to
eliminate the first-order term in Eq.~(\ref{perturbation_expansion}). The bath correlation function $C(t,t')$ is assumed to be related to the 1/f spectrum. Returning to the original rotating frame, the time-nonlocal non-Markovian master equation for the reduced system density matrix $\rho(t)$ takes the form
\begin{equation}
\label{master_eq_solved}
\dot{\rho}(t)=-\frac{i}{\hbar}[H_S(t),\rho(t)] + \{[\mathcal{K}_c(t),\sigma_z]+[\sigma_z,\mathcal{K}_c^\dagger(t)]\},   
\end{equation} 
where $H_{\text{S}}(t) = [\hbar \delta(t)/2]\, \sigma_z+[\hbar \Omega(t)/2]\, \sigma_x$ is the Hamiltonian engineering part of the system, and the dissipation kernel takes the form of
\begin{equation}
\label{kernel_tobe_transformed}
\mathcal{K}_c(t) = U_S^{\dagger}(t)\left[\int^t_0 dt' C(t,t')\tilde{\sigma_z}(t')\tilde{\rho}(t')\right] U_S(t),
\end{equation}
with $\tilde{\rho}(t)=U_S(t)\rho(t)U_S^\dagger(t)$.
We can further rewrite Eq.~(\ref{kernel_tobe_transformed}) in a superoperator form as
\begin{equation}
\label{transformed_kernel}
\mathcal{K}_c(t) = \int^t_0 dt'C(t,t') \mathcal{U}_S(t,t') \sigma_z\rho(t').
\end{equation}
Here we have the superoperator of the unitary qubit system propagator $\mathcal{U}_S(t,t') A= T_+ e^{-\frac{i}{\hbar}\int^{t}_{t'}d \tau [H_S, A] }$ which propagates $A$ following the system Hamiltonian with $T_+$ being the time-ordering operator. However, there are several methods proving that the Born approximation is equivalent to the time-local approximation when we expand the Born series and Markovian series \cite{PhysRevB.90.104302}. This will give an analytic form of the master equation without convolution but still preserves the non-Markovian effect, and the detailed simplifying process was derived from Refs. \cite{PhysRevB.90.104302, Zou2024, 10.1093/acprof:oso/9780199213900.001.0001}.  Starting from the well-known Nakajima-Zwanzig equation, which is a time-nonlocal equation, the superoperator action \ref{transformed_kernel} when the coupling is sufficiently weak can be truncated up to the second order and yield
\begin{equation}
\label{tidy_up_kernel}
    \mathcal{K}(t)\rho(t) \equiv \mathcal{K}_c(t) = \int^t_0 dt'C(t,t') \mathcal{U}_S(t,t') \sigma_z\rho(t).
\end{equation}
Therefore, we rewrite the master equation using a time-convolutionless (time-local) superoperator and obtain Eqs. ~(\ref{master_eq}) and (\ref{dissipation_kernel}) in the main text.

\section{\label{appendix D} The dynamics of the off-diagonal term}

It is also obvious that the non-Markovian effect can be seen by solving the master equation without driving or other detuning. For pure dephasing without control, Eq.~(\ref{dephase_no_control}) provides an analytic result. For the spectrum of the correlation function Eq.~(\ref{OU_corr}), the distribution of $\gamma_i$ is usually chosen to be $P(\gamma_i) \propto 1/\gamma_i$ for a $1/f$ noise spectrum in the numerical fitting \cite{Schriefl_2006, RevModPhys.53.497, RevModPhys.86.361, physics/0204033}. Then replace the correlations in Eq.~(\ref{dephase_no_control})  with the numerical form of Eq.~(\ref{numerical_corr_general}) with parameters given by Eqs.~(\ref{theta_i} and (\ref{C_i}) for the $1/f$ noise. Then the master equation can be written as
\begin{align}
\label{dephasing_no_control_numerical}
\frac{d\rho_{01}(t)}{dt} = \int^t_0 \sum_j C_i(0) e^{-\gamma_i|t-t'|}  dt' \rho_{01}(t).
\end{align}
For the integral part on the right-hand side of Eq.~(\ref{dephasing_no_control_numerical}),
\begin{align}
    \int^t_0 dt \frac{\sigma_i^2}{\gamma_i^2}e^{-\gamma_i|t-t'|} = \frac{\sigma_i^2}{\gamma_i^3}(1-e^{-\gamma_it}).
\end{align}
The differential equation of the off-diagonal term can be simplified to 
\begin{align}
    \frac{d \rho_{01}}{dt} = -\sum_i \frac{\sigma_i^2}{\gamma_i^3}(1-e^{-\gamma_it}) \rho_{01}.
\end{align}
With the initial condition $\rho_{01}=0.5$ when the qubit is in the azimuthal plane, the off-diagonal element can be obtained in analytic form as 
\begin{align}
\label{off_diagonal_decay}
    \rho_{01} = \frac{1}{2}e^{-\tilde{G}(t)},
\end{align}
where
\begin{align}
    \tilde{G}(t) = \sum_i \frac{\sigma_i^2}{\gamma_i^3} \left( t+\frac{1}{\gamma_i}e^{-\gamma_i t}-\frac{1}{\gamma_i} \right).
\end{align}
 By expanding the series of $\tilde{G}(t)$ to the second order of $\gamma_i t$, we then get
\begin{align}
    \tilde{G}(t) &\approx \sum_i \frac{\sigma_i^2}{\gamma_i^3}\left(t-\frac{1}{\gamma_i}+\frac{1}{\gamma_i}(1-\gamma_it+\gamma_i^2t^2)\right)\\
    &=\sum_i \frac{\sigma_i^2}{\gamma_i^2}t^2.
\end{align}
The off-diagonal element of Eq.~(\ref{off_diagonal_decay}) is therefore dominated by 
\begin{align}
    \rho_{01} =\frac{1}{2} e^{-(\frac{t}{T_2})^2},
\end{align}
following the theoretical prediction.

\section{\label{appendix E} Analyzing GST estimates}

Assume that the log-likelihood ratio statistic between the GST estimate and the maximal likelihood model is a $\chi_k^2$ random variable \cite{Wilks1938, Nielsen2021}:
\begin{align}
    2(\log\mathcal{L}_{max}-\log\mathcal{L}) \approx \chi_k^2,
\end{align}
where the maximal likelihood model has $k$ more parameters than the gate set model has non-gauge parameters. The $\chi_k^2$ distribution has mean $k$ and standard deviation $\sqrt{2k}$. The observed model violation is quantified by the number of standard deviations by which the log-likelihood ratio exceeds its expected value under the $\chi_k^2$ hypothesis:
\begin{align}
    N_\sigma = \frac{2(\log\mathcal{L}_{max}-\log\mathcal{L})-k}{2\sqrt{k}}.
\end{align}
If the observed log-likelihood ratio is too high to be sampled from a $\chi_k^2$ distribution, then we have evidence that the data were influenced by a non-Markovian process. To be more specific, $N_{\sigma} \ll 1$ indicates an extremely good fit that appears completely trustworthy. In contrast, $N_{\sigma}  \gg 1$ indicates a significant model violation, which means that no gate set can describe all of the data. Model violation indicates the presence of some kind of non-Markovian noise.


\begin{thebibliography}{10}

\bibitem{PhysRevB.72.134519}
G.~Ithier, E.~Collin, P.~Joyez, P.~J. Meeson, D.~Vion, D.~Esteve, F.~Chiarello, A.~Shnirman, Y.~Makhlin, J.~Schriefl, and G.~Sch\"on.
\newblock Decoherence in a superconducting quantum bit circuit.
\newblock {\em Physical Review B}, 72:134519, Oct 2005.

\bibitem{Schriefl_2006}
Josef Schriefl, Yuriy Makhlin, Alexander Shnirman, and Gerd Schön.
\newblock Decoherence from ensembles of two-level fluctuators.
\newblock {\em New Journal of Physics}, 8(1):1, jan 2006.

\bibitem{RevModPhys.88.021002}
Heinz-Peter Breuer, Elsi-Mari Laine, Jyrki Piilo, and Bassano Vacchini.
\newblock Colloquium: Non-markovian dynamics in open quantum systems.
\newblock {\em Rev. Mod. Phys.}, 88:021002, Apr 2016.

\bibitem{Nielsen2021}
Erik Nielsen, John~King Gamble, Kenneth Rudinger, Travis Scholten, Kevin Young, and Robin Blume-Kohout.
\newblock Gate set tomography.
\newblock {\em Quantum}, 5:557, October 2021.

\bibitem{Hashim2023}
Akel Hashim, Stefan Seritan, Timothy Proctor, Kenneth Rudinger, Noah Goss, Ravi~K. Naik, John~Mark Kreikebaum, David~I. Santiago, and Irfan Siddiqi.
\newblock Benchmarking quantum logic operations relative to thresholds for fault tolerance.
\newblock {\em npj Quantum Information}, 9(1), October 2023.

\bibitem{White2020}
G.~A.~L. White, C.~D. Hill, F.~A. Pollock, L.~C.~L. Hollenberg, and K.~Modi.
\newblock Demonstration of non-markovian process characterisation and control on a quantum processor.
\newblock {\em Nature Communications}, 11(1), December 2020.

\bibitem{arxiv.2307.12452}
R.~Y. Su, J.~Y. Huang, N.~Dumoulin. Stuyck, M.~K. Feng, W.~Gilbert, T.~J. Evans, W.~H. Lim, F.~E. Hudson, K.~W. Chan, W.~Huang, Kohei~M. Itoh, R.~Harper, S.~D. Bartlett, C.~H. Yang, A.~Laucht, A.~Saraiva, T.~Tanttu, and A.~S. Dzurak.
\newblock Characterizing non-markovian quantum process by fast bayesian tomography, 2023.

\bibitem{Proctor2020}
Timothy Proctor, Melissa Revelle, Erik Nielsen, Kenneth Rudinger, Daniel Lobser, Peter Maunz, Robin Blume-Kohout, and Kevin Young.
\newblock Detecting and tracking drift in quantum information processors.
\newblock {\em Nature Communications}, 11(1), October 2020.

\bibitem{Mavadia2018}
S.~Mavadia, C.~L. Edmunds, C.~Hempel, H.~Ball, F.~Roy, T.~M. Stace, and M.~J. Biercuk.
\newblock Experimental quantum verification in the presence of temporally correlated noise.
\newblock {\em npj Quantum Information}, 4(1), February 2018.

\bibitem{Dehollain2016}
Juan~P Dehollain, Juha~T Muhonen, Robin Blume-Kohout, Kenneth~M Rudinger, John~King Gamble, Erik Nielsen, Arne Laucht, Stephanie Simmons, Rachpon Kalra, Andrew~S Dzurak, and Andrea Morello.
\newblock Optimization of a solid-state electron spin qubit using gate set tomography.
\newblock {\em New Journal of Physics}, 18(10):103018, October 2016.

\bibitem{li2024}
Ze-Tong Li, Cong-Cong Zheng, Fan-Xu Meng, Han Zeng, Tian Luan, Zai-Chen Zhang, and Xu-Tao Yu.
\newblock Non-markovian quantum gate set tomography.
\newblock {\em Quantum Science and Technology}, 9(3):035027, May 2024.

\bibitem{Yoneda2017}
Jun Yoneda, Kenta Takeda, Tomohiro Otsuka, Takashi Nakajima, Matthieu~R. Delbecq, Giles Allison, Takumu Honda, Tetsuo Kodera, Shunri Oda, Yusuke Hoshi, Noritaka Usami, Kohei~M. Itoh, and Seigo Tarucha.
\newblock A quantum-dot spin qubit with coherence limited by charge noise and fidelity higher than 99.9\%.
\newblock {\em Nature Nanotechnology}, 13(2):102--106, December 2017.

\bibitem{PhysRevB.77.174509}
\L{}ukasz Cywi\ifmmode~\acute{n}\else \'{n}\fi{}ski, Roman~M. Lutchyn, Cody~P. Nave, and S.~Das~Sarma.
\newblock How to enhance dephasing time in superconducting qubits.
\newblock {\em Physical Review B}, 77:174509, May 2008.

\bibitem{Yoneda2023}
J.~Yoneda, J.~S. Rojas-Arias, P.~Stano, K.~Takeda, A.~Noiri, T.~Nakajima, D.~Loss, and S.~Tarucha.
\newblock Noise-correlation spectrum for a pair of spin qubits in silicon.
\newblock {\em Nature Physics}, 19(12):1793--1798, October 2023.

\bibitem{arxiv.2309.12542}
Amanda~E. Seedhouse, Nard~Dumoulin Stuyck, Santiago Serrano, Tuomo Tanttu, Will Gilbert, Jonathan~Yue Huang, Fay~E. Hudson, Kohei~M. Itoh, Arne Laucht, Wee~Han Lim, Chih~Hwan Yang, Andrew~S. Dzurak, and Andre Saraiva.
\newblock Spatio-temporal correlations of noise in mos spin qubits, 2023.

\bibitem{Keith2022}
D.~Keith, S.~K. Gorman, Y.~He, L.~Kranz, and M.~Y. Simmons.
\newblock Impact of charge noise on electron exchange interactions in semiconductors.
\newblock {\em npj Quantum Information}, 8(1), February 2022.

\bibitem{Dial2016}
Oliver Dial, Douglas~T McClure, Stefano Poletto, G~A Keefe, Mary~Beth Rothwell, Jay~M Gambetta, David~W Abraham, Jerry~M Chow, and Matthias Steffen.
\newblock Bulk and surface loss in superconducting transmon qubits.
\newblock {\em Superconductor Science and Technology}, 29(4):044001, March 2016.

\bibitem{RevModPhys.95.025003}
Guido Burkard, Thaddeus~D. Ladd, Andrew Pan, John~M. Nichol, and Jason~R. Petta.
\newblock Semiconductor spin qubits.
\newblock {\em Rev. Mod. Phys.}, 95:025003, Jun 2023.

\bibitem{PhysRevLett.88.186802}
Alexander~V. Khaetskii, Daniel Loss, and Leonid Glazman.
\newblock Electron spin decoherence in quantum dots due to interaction with nuclei.
\newblock {\em Physical Review Letters}, 88:186802, Apr 2002.

\bibitem{Xue2022}
Xiao Xue, Maximilian Russ, Nodar Samkharadze, Brennan Undseth, Amir Sammak, Giordano Scappucci, and Lieven M.~K. Vandersypen.
\newblock Quantum logic with spin qubits crossing the surface code threshold.
\newblock {\em Nature}, 601(7893):343--347, January 2022.

\bibitem{PhysRevLett.96.047202}
Yasuhiro Tokura, Wilfred~G. van~der Wiel, Toshiaki Obata, and Seigo Tarucha.
\newblock Coherent single electron spin control in a slanting zeeman field.
\newblock {\em Physical Review Letters}, 96:047202, Jan 2006.

\bibitem{PioroLadrire2008}
M.~Pioro-Ladrière, T.~Obata, Y.~Tokura, Y.-S. Shin, T.~Kubo, K.~Yoshida, T.~Taniyama, and S.~Tarucha.
\newblock Electrically driven single-electron spin resonance in a slanting zeeman field.
\newblock {\em Nature Physics}, 4(10):776--779, August 2008.

\bibitem{PhysRevB.74.165319}
Vitaly~N. Golovach, Massoud Borhani, and Daniel Loss.
\newblock Electric-dipole-induced spin resonance in quantum dots.
\newblock {\em Physical Review B}, 74:165319, Oct 2006.

\bibitem{PhysRevB.97.085421}
Maximilian Russ, D.~M. Zajac, A.~J. Sigillito, F.~Borjans, J.~M. Taylor, J.~R. Petta, and Guido Burkard.
\newblock High-fidelity quantum gates in si/sige double quantum dots.
\newblock {\em Physical Review B}, 97:085421, Feb 2018.

\bibitem{PioroLadrire2007}
M.~Pioro-Ladrière, Y.~Tokura, T.~Obata, T.~Kubo, and S.~Tarucha.
\newblock Micromagnets for coherent control of spin-charge qubit in lateral quantum dots.
\newblock {\em Applied Physics Letters}, 90(2), January 2007.

\bibitem{Takeda2016}
Kenta Takeda, Jun Kamioka, Tomohiro Otsuka, Jun Yoneda, Takashi Nakajima, Matthieu~R. Delbecq, Shinichi Amaha, Giles Allison, Tetsuo Kodera, Shunri Oda, and Seigo Tarucha.
\newblock A fault-tolerant addressable spin qubit in a natural silicon quantum dot.
\newblock {\em Science Advances}, 2(8), August 2016.

\bibitem{Huang2024}
Jonathan~Y. Huang, Rocky~Y. Su, Wee~Han Lim, MengKe Feng, Barnaby van Straaten, Brandon Severin, Will Gilbert, Nard Dumoulin~Stuyck, Tuomo Tanttu, Santiago Serrano, Jesus~D. Cifuentes, Ingvild Hansen, Amanda~E. Seedhouse, Ensar Vahapoglu, Ross C.~C. Leon, Nikolay~V. Abrosimov, Hans-Joachim Pohl, Michael L.~W. Thewalt, Fay~E. Hudson, Christopher~C. Escott, Natalia Ares, Stephen~D. Bartlett, Andrea Morello, Andre Saraiva, Arne Laucht, Andrew~S. Dzurak, and Chih~Hwan Yang.
\newblock High-fidelity spin qubit operation and algorithmic initialization above 1 k.
\newblock {\em Nature}, 627(8005):772--777, March 2024.

\bibitem{Noiri2022-xw}
Akito Noiri, Kenta Takeda, Takashi Nakajima, Takashi Kobayashi, Amir Sammak, Giordano Scappucci, and Seigo Tarucha.
\newblock Fast universal quantum gate above the fault-tolerance threshold in silicon.
\newblock {\em Nature}, 601(7893):338--342, January 2022.

\bibitem{Mills2022}
Adam~R. Mills, Charles~R. Guinn, Michael~J. Gullans, Anthony~J. Sigillito, Mayer~M. Feldman, Erik Nielsen, and Jason~R. Petta.
\newblock Two-qubit silicon quantum processor with operation fidelity exceeding 99\%.
\newblock {\em Science Advances}, 8(14), April 2022.

\bibitem{doi:10.1126/science.aao5965}
D.~M. Zajac, A.~J. Sigillito, M.~Russ, F.~Borjans, J.~M. Taylor, G.~Burkard, and J.~R. Petta.
\newblock Resonantly driven cnot gate for electron spins.
\newblock {\em Science}, 359(6374):439--442, 2018.

\bibitem{PhysRevB.77.180502}
J.~A. Schreier, A.~A. Houck, Jens Koch, D.~I. Schuster, B.~R. Johnson, J.~M. Chow, J.~M. Gambetta, J.~Majer, L.~Frunzio, M.~H. Devoret, S.~M. Girvin, and R.~J. Schoelkopf.
\newblock Suppressing charge noise decoherence in superconducting charge qubits.
\newblock {\em Phys. Rev. B}, 77:180502, May 2008.

\bibitem{PhysRevA.99.042310}
Chia-Hsien Huang, Chih-Hwan Yang, Chien-Chang Chen, Andrew~S. Dzurak, and Hsi-Sheng Goan.
\newblock High-fidelity and robust two-qubit gates for quantum-dot spin qubits in silicon.
\newblock {\em Physical Review A}, 99:042310, Apr 2019.

\bibitem{PhysRevApplied.17.024068}
T.J. Evans, W.~Huang, J.~Yoneda, R.~Harper, T.~Tanttu, K.W. Chan, F.E. Hudson, K.M. Itoh, A.~Saraiva, C.H. Yang, A.S. Dzurak, and S.D. Bartlett.
\newblock Fast bayesian tomography of a two-qubit gate set in silicon.
\newblock {\em Physical Review Applied}, 17:024068, Feb 2022.

\bibitem{PRXQuantum.5.010306}
M.J. Gullans, M.~Caranti, A.R. Mills, and J.R. Petta.
\newblock Compressed gate characterization for quantum devices with time-correlated noise.
\newblock {\em Physical Review X}, 5:010306, Jan 2024.

\bibitem{Krotov1993}
V.~F. Krotov.
\newblock Global methods in optimal control theory.
\newblock In {\em Advances in Nonlinear Dynamics and Control: A Report from Russia}, page 74–121. Birkh\"{a}user, Boston, 1993.

\bibitem{Krotov1996}
Vadim~F. Krotov.
\newblock {\em Global Methods in Optimal Control Theory (Monographs and Textbooks in Pure and Applied Mathematics)}, volume 195.
\newblock Marcel Dekker, New York, Feburary 1996.

\bibitem{Yang2019-yk}
C~H Yang, K~W Chan, R~Harper, W~Huang, T~Evans, J~C~C Hwang, B~Hensen, A~Laucht, T~Tanttu, F~E Hudson, S~T Flammia, K~M Itoh, A~Morello, S~D Bartlett, and A~S Dzurak.
\newblock Silicon qubit fidelities approaching incoherent noise limits via pulse engineering.
\newblock {\em Nature Electronics}, 2(4):151--158, April 2019.

\bibitem{Zou2024}
Ji~Zou, Stefano Bosco, and Daniel Loss.
\newblock Spatially correlated classical and quantum noise in driven qubits.
\newblock {\em npj Quantum Information}, 10(1), April 2024.

\bibitem{PhysRevB.90.104302}
Christian Karlewski and Michael Marthaler.
\newblock Time-local master equation connecting the born and markov approximations.
\newblock {\em Physical Review B}, 90:104302, Sep 2014.

\bibitem{10.1093/acprof:oso/9780199213900.001.0001}
Heinz-Peter Breuer and Francesco Petruccione.
\newblock {\em {The Theory of Open Quantum Systems}}.
\newblock Oxford University Press, 01 2007.

\bibitem{PhysRevA.95.022121}
Gerardo~A. Paz-Silva, Leigh~M. Norris, and Lorenza Viola.
\newblock Multiqubit spectroscopy of gaussian quantum noise.
\newblock {\em Physical Review A}, 95:022121, Feb 2017.

\bibitem{PhysRevA.86.042107}
Daniel Maldonado-Mundo, Patrik \"Ohberg, Brendon~W. Lovett, and Erika Andersson.
\newblock Investigating the generality of time-local master equations.
\newblock {\em Physical Review A}, 86:042107, Oct 2012.

\bibitem{RevModPhys.53.497}
P.~Dutta and P.~M. Horn.
\newblock Low-frequency fluctuations in solids: $\frac{1}{f}$ noise.
\newblock {\em Rev. Mod. Phys.}, 53:497--516, Jul 1981.

\bibitem{RevModPhys.86.361}
E.~Paladino, Y.~M. Galperin, G.~Falci, and B.~L. Altshuler.
\newblock 1/f noise: Implications for solid-state quantum information.
\newblock {\em Reviews of Modern Physics}, 86:361--418, Apr 2014.

\bibitem{physics/0204033}
Edoardo Milotti.
\newblock 1/f noise: a pedagogical review, 2002.

\bibitem{Tannor1992}
D.~J. Tannor, V.~A. Kazakov, and V.~Orlov.
\newblock Control of photochemical branching: Novel procedures for finding optimal pulses and global upper bounds.
\newblock In J.~Broeckhove and L.~Lathouwers, editors, {\em Time-Dependent Quantum Molecular Dynamics}, volume 299 of {\em NATO Advanced Studies Institute, Series B: Physics}, page 347–360. Springer New York, 1992.

\bibitem{PhysRevLett.89.188301}
Jos\'e~P. Palao and Ronnie Kosloff.
\newblock Quantum computing by an optimal control algorithm for unitary transformations.
\newblock {\em Phys. Rev. Lett.}, 89:188301, Oct 2002.

\bibitem{PhysRevA.68.062308}
Jos\'e~P. Palao and Ronnie Kosloff.
\newblock Optimal control theory for unitary transformations.
\newblock {\em Phys. Rev. A}, 68:062308, Dec 2003.

\bibitem{Jirari2009}
H.~Jirari.
\newblock Optimal control approach to dynamical suppression of decoherence of a qubit.
\newblock {\em EPL (Europhysics Letters)}, 87(4):40003, August 2009.

\bibitem{Wenin2009}
M.~Wenin, R.~Roloff, and W.~P\"{o}tz.
\newblock Robust control of josephson charge qubits.
\newblock {\em Journal of Applied Physics}, 105(8), April 2009.

\bibitem{PhysRevB.79.224516}
Robert Roloff and Walter P\"otz.
\newblock Time-optimal performance of josephson charge qubits: A process tomography approach.
\newblock {\em Phys. Rev. B}, 79:224516, Jun 2009.

\bibitem{9872062}
Siyuan Niu and Aida Todri-Sanial.
\newblock Effects of dynamical decoupling and pulse-level optimizations on ibm quantum computers.
\newblock {\em IEEE Transactions on Quantum Engineering}, 3:1--10, 2022.

\bibitem{PhysRevA.83.053426}
Reuven Eitan, Michael Mundt, and David~J. Tannor.
\newblock Optimal control with accelerated convergence: Combining the krotov and quasi-newton methods.
\newblock {\em Physical Review A}, 83:053426, May 2011.

\bibitem{Maximov2008}
Ivan~I. Maximov, Zdenĕk Tošner, and Niels~Chr. Nielsen.
\newblock Optimal control design of nmr and dynamic nuclear polarization experiments using monotonically convergent algorithms.
\newblock {\em The Journal of Chemical Physics}, 128(18), May 2008.

\bibitem{PhysRevA.85.032321}
Bin Hwang and Hsi-Sheng Goan.
\newblock Optimal control for non-markovian open quantum systems.
\newblock {\em Physical Review A}, 85:032321, Mar 2012.

\bibitem{PhysRevA.89.062310}
Jung-Shen Tai, Kuan-Ting Lin, and Hsi-Sheng Goan.
\newblock Optimal control of quantum gates in an exactly solvable non-markovian open quantum bit system.
\newblock {\em Phys. Rev. A}, 89:062310, Jun 2014.

\bibitem{PhysRevA.90.012318}
Shang-Yu Huang and Hsi-Sheng Goan.
\newblock Optimal control for fast and high-fidelity quantum gates in coupled superconducting flux qubits.
\newblock {\em Phys. Rev. A}, 90:012318, Jul 2014.

\bibitem{PhysRevA.91.052315}
Yi~Chou, Shang-Yu Huang, and Hsi-Sheng Goan.
\newblock Optimal control of fast and high-fidelity quantum gates with electron and nuclear spins of a nitrogen-vacancy center in diamond.
\newblock {\em Phys. Rev. A}, 91:052315, May 2015.

\bibitem{PhysRevLett.102.090401}
P.~Rebentrost, I.~Serban, T.~Schulte-Herbr\"uggen, and F.~K. Wilhelm.
\newblock Optimal control of a qubit coupled to a non-markovian environment.
\newblock {\em Phys. Rev. Lett.}, 102:090401, Mar 2009.

\bibitem{10.21468/SciPostPhys.7.6.080}
Michael~H. Goerz, Daniel Basilewitsch, Fernando Gago-Encinas, Matthias~G. Krauss, Karl~P. Horn, Daniel~M. Reich, and Christiane~P. Koch.
\newblock {Krotov: A Python implementation of Krotov's method for quantum optimal control}.
\newblock {\em SciPost Phys.}, 7:080, 2019.

\bibitem{PhysRevB.102.035306}
Chien-Chang Chen, Thomas~M. Stace, and Hsi-Sheng Goan.
\newblock Full-polaron master equation approach to dynamical steady states of a driven two-level system beyond the weak system-environment coupling.
\newblock {\em Phys. Rev. B}, 102:035306, Jul 2020.

\bibitem{Goerz2014-wa}
Michael~H Goerz, Daniel~M Reich, and Christiane~P Koch.
\newblock Optimal control theory for a unitary operation under dissipative evolution.
\newblock {\em New J. Phys.}, 16(5):055012, May 2014.

\bibitem{Bartana1993-za}
Allon Bartana, Ronnie Kosloff, and David~J Tannor.
\newblock Laser cooling of molecular internal degrees of freedom by a series of shaped pulses.
\newblock {\em J. Chem. Phys.}, 99(1):196--210, July 1993.

\bibitem{Ohtsuki1999-du}
Yukiyoshi Ohtsuki, Wusheng Zhu, and Herschel Rabitz.
\newblock Monotonically convergent algorithm for quantum optimal control with dissipation.
\newblock {\em J. Chem. Phys.}, 110(20):9825--9832, May 1999.

\bibitem{PhysRevA.93.032113}
Chien-Chang Chen and Hsi-Sheng Goan.
\newblock Effects of initial system-environment correlations on open-quantum-system dynamics and state preparation.
\newblock {\em Phys. Rev. A}, 93:032113, Mar 2016.

\bibitem{Yang2016}
Xu-Chen Yang and Xin Wang.
\newblock Noise filtering of composite pulses for singlet-triplet qubits.
\newblock {\em Scientific Reports}, 6(1), July 2016.

\bibitem{Stehouwer2025}
Lucas E.~A. Stehouwer, Cécile~X. Yu, Barnaby van Straaten, Alberto Tosato, Valentin John, Davide Degli~Esposti, Asser Elsayed, Davide Costa, Stefan~D. Oosterhout, Nico~W. Hendrickx, Menno Veldhorst, Francesco Borsoi, and Giordano Scappucci.
\newblock Exploiting strained epitaxial germanium for scaling low-noise spin qubits at the micrometre scale.
\newblock {\em Nature Materials}, 24(12):1906–1912, 2025.

\bibitem{PhysRevB.111.155403}
Zhanning Wang, Sina Gholizadeh, Xuedong Hu, S.~Das~Sarma, and Dimitrie Culcer.
\newblock Dephasing of planar ge hole spin qubits due to $1/f$ charge noise.
\newblock {\em Phys. Rev. B}, 111:155403, Apr 2025.

\bibitem{arxiv.2306.06229}
Joel~B. Varley, Keith~G. Ray, and Vincenzo Lordi.
\newblock Dangling bonds as possible contributors to charge noise in silicon and silicon-germanium quantum dot qubits, 2023.

\bibitem{Elsayed2024}
A.~Elsayed, M.~M.~K. Shehata, C.~Godfrin, S.~Kubicek, S.~Massar, Y.~Canvel, J.~Jussot, G.~Simion, M.~Mongillo, D.~Wan, B.~Govoreanu, I.~P. Radu, R.~Li, P.~Van~Dorpe, and K.~De~Greve.
\newblock Low charge noise quantum dots with industrial cmos manufacturing.
\newblock {\em npj Quantum Information}, 10(1), 2024.

\bibitem{Culcer2009}
Dimitrie Culcer, Xuedong Hu, and S.~Das~Sarma.
\newblock Dephasing of si spin qubits due to charge noise.
\newblock {\em Applied Physics Letters}, 95(7), August 2009.

\bibitem{Steinacker2025}
Paul Steinacker, Nard Dumoulin~Stuyck, Wee~Han Lim, Tuomo Tanttu, MengKe Feng, Santiago Serrano, Andreas Nickl, Marco Candido, Jesus~D. Cifuentes, Ensar Vahapoglu, Samuel~K. Bartee, Fay~E. Hudson, Kok~Wai Chan, Stefan Kubicek, Julien Jussot, Yann Canvel, Sofie Beyne, Yosuke Shimura, Roger Loo, Clement Godfrin, Bart Raes, Sylvain Baudot, Danny Wan, Arne Laucht, Chih~Hwan Yang, Andre Saraiva, Christopher~C. Escott, Kristiaan De~Greve, and Andrew~S. Dzurak.
\newblock Industry-compatible silicon spin-qubit unit cells exceeding 99
\newblock {\em Nature}, 646(8083):81–87, 2025.

\bibitem{https://doi.org/10.11578/dc.20190722.2}
Erik Nielsen, Robin Blume-Kohout, Kenneth Rudinger, Timothy Proctor, and Lucas Saldyt.
\newblock Python gst implementation (pygsti) v. 0.9, 2019.

\bibitem{Mdzik2022}
Mateusz~T. Madzik, Serwan Asaad, Akram Youssry, Benjamin Joecker, Kenneth~M. Rudinger, Erik Nielsen, Kevin~C. Young, Timothy~J. Proctor, Andrew~D. Baczewski, Arne Laucht, Vivien Schmitt, Fay~E. Hudson, Kohei~M. Itoh, Alexander~M. Jakob, Brett~C. Johnson, David~N. Jamieson, Andrew~S. Dzurak, Christopher Ferrie, Robin Blume-Kohout, and Andrea Morello.
\newblock Precision tomography of a three-qubit donor quantum processor in silicon.
\newblock {\em Nature}, 601(7893):348--353, January 2022.

\bibitem{arxiv.2410.15590}
Paul Steinacker, Nard~Dumoulin Stuyck, Wee~Han Lim, Tuomo Tanttu, MengKe Feng, Andreas Nickl, Santiago Serrano, Marco Candido, Jesus~D. Cifuentes, Fay~E. Hudson, Kok~Wai Chan, Stefan Kubicek, Julien Jussot, Yann Canvel, Sofie Beyne, Yosuke Shimura, Roger Loo, Clement Godfrin, Bart Raes, Sylvain Baudot, Danny Wan, Arne Laucht, Chih~Hwan Yang, Andre Saraiva, Christopher~C. Escott, Kristiaan De~Greve, and Andrew~S. Dzurak.
\newblock A 300 mm foundry silicon spin qubit unit cell exceeding 99\% fidelity in all operations, 2024.

\bibitem{PhysRevA.96.022330}
David~C. McKay, Christopher~J. Wood, Sarah Sheldon, Jerry~M. Chow, and Jay~M. Gambetta.
\newblock Efficient $z$ gates for quantum computing.
\newblock {\em Physical Review A}, 96:022330, Aug 2017.

\bibitem{Wang2024}
Chien-An Wang, Valentin John, Hanifa Tidjani, Cécile~X. Yu, Alexander~S. Ivlev, Corentin Déprez, Floor van Riggelen-Doelman, Benjamin~D. Woods, Nico~W. Hendrickx, William I.~L. Lawrie, Lucas E.~A. Stehouwer, Stefan~D. Oosterhout, Amir Sammak, Mark Friesen, Giordano Scappucci, Sander~L. de~Snoo, Maximilian Rimbach-Russ, Francesco Borsoi, and Menno Veldhorst.
\newblock Operating semiconductor quantum processors with hopping spins.
\newblock {\em Science}, 385(6707):447--452, July 2024.

\bibitem{vandenBerg2023}
Ewout van~den Berg, Zlatko~K. Minev, Abhinav Kandala, and Kristan Temme.
\newblock Probabilistic error cancellation with sparse pauli-lindblad models on noisy quantum processors.
\newblock {\em Nature Physics}, 19(8):1116--1121, May 2023.

\bibitem{Green2013}
Todd~J Green, Jarrah Sastrawan, Hermann Uys, and Michael~J Biercuk.
\newblock Arbitrary quantum control of qubits in the presence of universal noise.
\newblock {\em New Journal of Physics}, 15(9):095004, September 2013.

\bibitem{PRXQuantum.3.020335}
Robin Blume-Kohout, Marcus~P. da~Silva, Erik Nielsen, Timothy Proctor, Kenneth Rudinger, Mohan Sarovar, and Kevin Young.
\newblock A taxonomy of small markovian errors.
\newblock {\em Physical Review X}, 3:020335, May 2022.

\bibitem{PhysRevA.60.1888}
Micha\l{} Horodecki, Pawe\l{} Horodecki, and Ryszard Horodecki.
\newblock General teleportation channel, singlet fraction, and quasidistillation.
\newblock {\em Physical Review A}, 60:1888--1898, Sep 1999.

\bibitem{Wilks1938}
S.~S. Wilks.
\newblock The large-sample distribution of the likelihood ratio for testing composite hypotheses.
\newblock {\em The Annals of Mathematical Statistics}, 9(1):60--62, March 1938.

\end{thebibliography}
\end{document}